\documentclass[aps,twocolumn,10pt,prc,floatfix,showpacs,preprintnumbers,amsmath,amssymb,nofootinbib,superscriptaddress]{revtex4-1}

\usepackage[normalem]{ulem}
\usepackage{wasysym}
\usepackage{color}
\usepackage{graphicx}
\usepackage{subfig}
\usepackage{dcolumn}    % Align table columns on decimal point
\usepackage{multirow, booktabs }
\usepackage{comment}
\usepackage{soul}

\usepackage[format=plain,labelsep=period]{caption}
\usepackage{physics, mathtools}   %, mathabx}
\usepackage [ english ]{ babel }

\usepackage{bigstrut}
\usepackage{amsmath,amsfonts,amsthm,bm}
\usepackage{CJK}
\usepackage[pdfstartview=FitH,
            CJKbookmarks=true,
            bookmarksnumbered=true,
            bookmarksopen=true,
            colorlinks,
            pdfborder=001,
            linkcolor=blue,
            anchorcolor=blue,
            citecolor=blue,
            urlcolor=blue,
            ]{hyperref}

\begin{document}

\newcommand{\varQ}{\mathbf{q}}
\newcommand{\varK}{\mathbf{k}}
\newcommand{\varX}{\mathbf{x}}
\newcommand{\varR}{\mathbf{r}}
\newcommand{\Sch}{ Schr\"{o}dinger }
\newcommand{\coeffSch}{-\frac{\hbar^2}{2m}}
\newcommand{\etal}{\textit{et al.} }
\newcommand{\secondfdv}[3]{ \frac{\delta^2 {#1}}{\delta #2 \, \delta #3} }

\newcommand{\FM}[1]{{\color{magenta} #1}}
\newcommand{\xrm}[1]{{\color{blue} #1}}
\newcommand{\enr}[1]{{\color{green} #1}}
\newcommand{\canenr}[1]{{\color{green}\sout{#1}}}
\newcommand{\can}[1]{{\color{red}\sout{#1}}}

%\title{Draft Static response}
%\title{Perturbed nuclear matter within a finite-number DFT framework}
%\title{DFT study of perturbed nuclear matter with a finite number of particles \\ \xrm{Density functional theory approach for perturbed nuclear matter\\ with a finite number of particles}  \\ \enr{Perturbed nuclear matter studied with \\Density Functional Theory with a finite number of particles}  }

\title{ Perturbed nuclear matter studied within \\Density Functional Theory with a finite number of particles}

\author{F. Marino}
\email{francesco.marino@unimi.it}
\affiliation{Dipartimento di Fisica ``Aldo Pontremoli'', Universit\`a degli Studi di Milano, 20133 Milano, Italy}
\affiliation{INFN,  Sezione di Milano, 20133 Milano, Italy}

\author{G.~Col\`{o}}
\affiliation{Dipartimento di Fisica ``Aldo Pontremoli'', Universit\`a degli Studi di Milano, 20133 Milano, Italy}
\affiliation{INFN,  Sezione di Milano, 20133 Milano, Italy}

\author{X. Roca-Maza}
\affiliation{Dipartimento di Fisica ``Aldo Pontremoli'', Universit\`a degli Studi di Milano, 20133 Milano, Italy}
\affiliation{INFN,  Sezione di Milano, 20133 Milano, Italy}

\author{E. Vigezzi}
\affiliation{INFN,  Sezione di Milano, 20133 Milano, Italy}

\begin{abstract}
    Nuclear matter is studied within the Density Functional Theory (DFT) framework. 
    Our method employs a finite number of nucleons in a box subject to periodic boundary conditions, in order to simulate infinite matter and study its response to an external static potential. 
    %allows to study nuclear matter perturbed by an external static potential. 
    We detail both the theoretical formalism and its computational implementation for pure neutron matter and symmetric nuclear matter with Skyrme-like Energy Density Functionals (EDFs). 
    The implementation of spin-orbit, in particular, is carefully discussed. 
    Our method is applied to the problem of the static response of nuclear matter and the impact of the perturbation on the energies, densities and level structure of the system is investigated.
    %The method is demonstrated by studying the static response function and probing its convergence to the thermodynamic limit. 
    Our work is a crucial step in our program of \textit{ab initio}-based nuclear EDFs [\href{https://journals.aps.org/prc/abstract/10.1103/PhysRevC.104.024315}{Phys. Rev. C 104, 024315 (2021)}] as it paves the way towards the goal of constraining the EDF surface terms on \textit{ab initio} calculations.
\end{abstract}

\maketitle

\section{Introduction}
\label{sec: intro}
Nuclear matter, an ideal infinite system made of strongly-interacting nucleons, is currently subject to intense study from multiple perspectives, due to its connections to the nuclear physics of finite nuclei \cite{rocamaza2018,Piekarewicz2019NeutronSkin,Burgio2020}, 
the astrophysics of neutron stars and gravitational waves \cite{HaenselNeutronStars,Burgio2021,Piekarewicz2022}, and the physics of cold Fermi gases \cite{BulgacUnitaryGas,Gandolfi2015}. 

Nuclear matter has been studied theoretically both within \textit{ab initio} theory and Density Functional Theory (DFT). 
In synthesis, \textit{ab initio} or first-principle methods aim at finding an exact or systematically improvable solution to the many-body problem starting from a Hamiltonian that describes the interactions among the constituent nucleons \cite{computational_nuclear,Hergert2020}.
DFT, on the other hand, 
%\can{adopts a more phenomenological, an effective description that}
maps the many-particle problem to a single-particle (s.p.) self-consistent (s.c.) problem that is based on the concept of an Energy Density Functional (EDF), i.e. on expressing the total energy of a generic system as a functional of its (generalized) densities \cite{Schunck2019,colo2020,Martin2020}. 
DFT is in principle an exact theory, but the EDF which are currently used rely heavily on phenomenology \cite{colo2020}. 
%\xrm{[COMMENT: EDF is based on an exact theory (DFT) while ab initio methods are not currently based on an exact theory since cannot be derived from QCD (chiral symmetry is not exact and not fully restored in nuclei). The real problem we have in DFT is the lack of a systematic improvable scheme.}

The equation of state (EOS), i.e. (at zero temperature) the energy per particle as a function of the neutron and proton densities, is the fundamental ground state (g.s.) property of homogeneous matter and has been the main target of most works, see the reviews Refs. \cite{OertelOES,rocamaza2018,Burgio2021}.
Another line of research has focused on inhomogeneous nuclear matter \cite{Gandolfi2015}, motivated by the fact that the inner crust of neutron stars is not uniform \cite{HaenselNeutronStars} and by the attractive possibility of constraining specific terms %(e.g. spin-orbit and tensor) 
of the nuclear EDFs (see e.g. Refs. \cite{Maris2013,Forbes2014,Rrapaj2016,Shen2019}).
%, e.g. for improving the description of neutron-rich systems.
Neutron and neutron-proton drops, i.e. nuclear matter confined by an external trap, have been studied e.g. in Refs. \cite{Pudliner1996,gandolfi2011,Maris2013,Boulet2018}. The problem of the response of nuclear matter subject to a weak periodic perturbation has also been tackled. 
The dynamical response function has been determined for rather general EDFs numerically \cite{Riz2020TimeDependent} and analytically (see Refs. \cite{PASTORE20151,Pastore2021} and references therein). Recently, Gezerlis and collaborators \cite{Gezerlis2016,Gezerlis2017,Gezerlis2021}, extending techniques used for the electron gas \cite{Moroni1995,SenatoreBook,Dornheim2017Density} and cold atoms \cite{Gandolfi2014Unitary}, have attacked the problem of the neutron matter static response \textit{ab initio} with the Auxiliary Field Diffusion Monte Carlo (AFDMC) method \cite{carlson2015,Tews2020}.

While the EOS and the static and dynamic response can be studied directly in the thermodynamic limit (TL) in the framework of DFT \cite{rocamaza2018,PASTORE20151}, most \textit{ab initio} methods simulate infinite matter by employing a finite number of particles (see e.g. Refs. \cite{computational_nuclear,LietzCompNucl,Hagen2014,Barbieri2017,Arthuis2022,Piarulli2020}). In fact, they are limited to few tens of fermions at most, which implies that \textit{ab initio} results are affected by finite-size (FS) effects. 
In this context, developing a finite-$\rm{A}$ DFT formalism for nuclear matter is important for two reasons. First, very large numbers of particles can be studied in DFT due to its low computational cost and thus a playground for understanding and handling FS effects is provided. In Refs. \cite{Gezerlis2021,Gezerlis2022Skyrme}, for example, \textit{ab initio} simulations of perturbed matter were extrapolated to the TL with the aid of DFT calculations.
Second, the finite-$\rm{A}$ DFT approach is instrumental in our program of constructing \textit{ab initio}-based EDFs started in Ref. \cite{Marino2021}, since it paves the way to matching \textit{ab initio} and DFT calculations with the same number of particles in a consistent manner.
%and the same boundary conditions in a consistent manner.
The EOS of uniform matter has already been employed in a local density approximation scheme \cite{Marino2021,Riz2020TimeDependent} to link the EDF to microscopic theory. 
Full-fledged EDFs, however, must incorporate surface terms that can act exclusively in non-uniform systems.
Perturbed nuclear matter, in this respect, is a promising candidate for setting constraints on the EDF surface contribution (see e.g. Ref. \cite{Dalfovo1995,Gandolfi2014Unitary,Gandolfi2015}). 

This work is devoted to a detailed description of the solution of the DFT problem for nuclear matter under the effect of an external perturbation for Skyrme-like EDFs. 
Our approach is based on simulating nuclear matter using a finite number of nucleons in a box on which periodic boundary conditions are imposed.
The formalism for pure neutron matter (PNM) and symmetric nuclear matter (SNM), together with its numerical implementation, are presented; a careful analysis of the treatment of spin-orbit is provided. 
The static response problem is then tackled with this method and the effect of the perturbation on the energies, densities and level structure of the system is investigated. 

%In this work, we present a detailed description of the solution of the DFT problem for pure neutron matter (PNM) and symmetric nuclear matter (SNM) under the effect of an external perturbation for Skyrme-like EDFs. In our approach, nuclear matter is described by a finite number of nucleons in a finite box subject to periodic boundary conditions. The formalism, including a careful analysis of spin-orbit terms, and the numerical implementation are presented, and  \FM{the code is made available.} We demonstrate our method by applying it to the static response problem and studying the FS effects and the convergence to the TL. 

This paper is structured as follows.
Section \ref{sec: dft formalism} is devoted to a detailed description to the finite-$\rm{A}$ nuclear DFT formalism and to its numerical implementation. 
%Section \ref{sec: dft formalism} is devoted to an overview of nuclear DFT (Sec. \ref{sec: overview dft}) and the description of infinite matter (Sec. \ref{sec: intro inf matter}) before delving into a detailed description of the finite-$\rm{A}$ DFT formalism (sec. \ref{sec: skyrme box}).
Section \ref{sec: static response theory} reviews the theory of the static response of homogeneous matter.
Results are presented in Sec. \ref{sec: results}.
Lastly, Section \ref{sec: conclusions} summarizes our work and presents future developments.

\section{Nuclear DFT formalism}

\label{sec: dft formalism}

\subsection{Overview of nuclear DFT}
\label{sec: overview dft}
We give a brief overview of nuclear DFT \cite{colo2020,Schunck2019}. Details are given in our previous work Ref. \cite{Marino2021} and references therein.

We consider quasi-local (or Skyrme-like) EDF models \cite{Schunck2019} for time-reversal-invariant systems, such as spin-saturated nuclear matter, and neglect pairing. 
We adopt the Kohn-Sham (KS) scheme \cite{Martin2020}, in which a representation in terms of s.p. orbitals $\psi_j(\varX)$ is introduced and the kinetic energy term is equal to that of a non-interacting Fermi system.
Then, the total energy of a generic system is written as a functional of number density $\rho_t(\varX)$, kinetic  density $\tau_t(\varX)$ and spin-orbit density $\mathbf{J}_t(\varX)$ (see App. \ref{app: dens orbitals} for their definition) with $t=0,1$ labelling isoscalar ($\rho_0=\rho_n+\rho_p$) and isovector ($\rho_1=\rho_n-\rho_p$) quantities, and has the following structure:
%\can{ The EDF for time-reversal-invariant systems, such as spin-saturated nuclear matter, and neglecting pairing is a functional of number density $\rho_t(\varX)$, kinetic  density $\tau_t(\varX)$ and spin-orbit density $\mathbf{J}_t(\varX)$ \xrm{--see Appendix for their definitions--} with $t=0,1$ labelling isoscalar ($\rho_0=\rho_n+\rho_p$) and isovector ($\rho_1=\rho_n-\rho_p$) quantities, and has the following structure:   }
\begin{equation}
\label{eq: edf basic structure}
    E = \int d\varX\, \mathcal{E}(\varX) = E_{kin} + E_{pot} + E_{ext}
\end{equation}
which comprises the kinetic energy, a nuclear potential energy term and possibly an external potential contribution,
\begin{align}
    &E_{kin} = \int d\varX\, \mathcal{E}_{kin}(\varX) = \int d\varX\, \frac{\hbar^2}{2m} \tau_0(\varX) , \\
    &E_{pot} = \int d\varR\, \mathcal{E}_{pot}(\varX), \\
    &E_{ext} = \sum_{t=0,1} \int d\varX\, \rho_t(\varX) v_{t}(\varX).
\end{align}
Throughout this work $\mathcal{E}_{pot}$ has the form \cite{Marino2021}
\begin{align}
    & \mathcal{E}_{pot}(\varX) =
    \sum_{t=0,1} \, \bigg(  
    \sum_\gamma \left( c_{\gamma,0} + c_{\gamma,1} \beta^2 \right) \rho_0^{\gamma+1}  \\
    & + C_t^{\tau}  \rho_t \tau_t + 
    C_t^{\Delta \rho} \rho_t \Delta \rho_t + C_{t}^{J} \mathbf{J}_{t}^2 + C_t^{\nabla J} \rho_t \nabla \cdot \mathbf{J}_t  \bigg) \nonumber
\end{align}
with $\beta=\rho_1/\rho_0$ being the isospin asymmetry.
The KS-DFT equations are found by minimizing the EDF w.r.t the s.p. orbitals $\psi_j^{*}(\varX)$ and read for protons and neutrons ($q=n,p$) \cite{Schunck2019}
%\can{ The DFT (or Kohn-Sham \cite{Martin2020}) equations are found by introducing a s.p. orbital representation and minimizing the EDF w.r.t the s.p. orbitals $\psi_j^{*}(\varX)$ and read (for $q=n,p$)    }
\begin{align}
\label{eq: skyrme HF eqs}
    \bigg[ 
    & - \nabla \cdot \frac{\hbar^2}{2m^*_q(\varX)} \nabla + U_q(\varX) + v_q(\varX) + \\
    & \mathbf{W}_q(\varX) \cdot 
    \left( -i\right) \left( \nabla \cross \mathbf{\sigma} \right)
    \bigg] 
    \psi_j(\varX) = \epsilon_j \psi_j(\varX) \nonumber
\end{align}
where the fields entering the equations are defined as
\begin{equation}
\label{eq: def mwan field}
    U_q = \fdv{E}{\rho_q} \qquad 
    \frac{\hbar^2}{2m_q^*} = \fdv{E}{\tau_q}
    \qquad \mathbf{W}_q = \fdv{E}{\mathbf{J}_q}.
\end{equation}
$m_q^{*}(\varX)$, $U_q(\varX)$ and $\mathbf{W}_q(\varX)$ are called effective mass, mean field and spin-orbit potential, respectively.

\subsection{Infinite nuclear matter}
\label{sec: intro inf matter}
Nuclear matter is an infinite system of nucleons that interact through the strong interaction only \cite{rocamaza2018}. (The Coulomb interaction is neglected.)
In the following we concentrate on zero-temperature and spin-unpolarized matter.
Moreover, we limit ourselves to the limiting cases of SNM ($\rho_n=\rho_p=\rho_0/2$) and PNM ($\rho_p$=0, $\rho_n=\rho_0$), although extensions are straightforward. 
The fundamental quantity that characterizes homogeneous matter is the EOS $e(\rho,\beta) = E(\rho,\beta)/A$, where $E$ is the total energy of the system and $e$ the energy per nucleon. 
We also remind that in homogeneous matter both the gradients of the density and the spin-orbit density vanish \cite{Marino2021}.

Some theoretical approaches attack nuclear matter directly in the TL. These include nuclear DFT \cite{PASTORE20151,rocamaza2018} and e.g. Self-consistent Green's functions \cite{Rios2020}.
Most \textit{ab initio} methods, though, simulate infinite matter by using a finite number of particles (see e.g. Refs. \cite{LietzCompNucl,Hagen2014,Piarulli2020}). Among them is AFDMC \cite{Tews2020}, that has been used extensively not only for the nuclear matter EOS, but also for inhomogeneous matter, namely neutron drops \cite{Maris2013}, as well as for  neutron matter response \cite{Gezerlis2017}.
DFT, too, can be formulated with a finite nucleon number, as proposed in Ref. \cite{Gezerlis2022Skyrme}.
The standard technique adopted in most studies \cite{LietzCompNucl,Tews2020} involves considering $\rm{A}$ fermions enclosed in a cubic box of size $L$ and volume $\Omega=L^3$ and imposing periodic boundary conditions (PBCs) on the wave function. The cell size is chosen such as the density of the system is a fixed and constant $\rho_0=\rm{A}/\Omega$.
In this framework, the TL corresponds to the limit in which both $\rm{A}$ and $L$ go to infinity while keeping $\rho_0$ fixed \cite{Fetter}.
The free gas (FG), that is the starting point for studying interacting matter, is described in terms of s.p. plane waves orbitals $e^{i\varK \cdot \varX}/\sqrt{\Omega}$ with wave number $\varK$ and kinetic energy $\frac{\hbar^2 \varK^2}{2m}$. As a consequence of PBCs, the momenta $\varK$ are quantized, i.e. $\varK = \frac{2\pi}{L} \mathbf{n}$ where $\mathbf{n}$ is a three-component vector of integer numbers. 
Since the energy depends on $\varK^2$ and thus on $\mathbf{n}^2$, a "momentum space" shell structure emerges, with different energy levels being labelled by $n^2$ and being degenerate. 
The first few momentum space "magic numbers" are given by $\rm{A}/g=$1, 7, 19, 27, 33 etc. \cite{LietzCompNucl}, where $g$ is spin/isospin degeneracy (2 for spin-saturated PNM, 4 for for spin-saturated SNM).
Typically, the number of fermions in a calculation is selected so as to correspond to a shell closure of the FG in both homogeneous and perturbed matter. As we discuss below, this choice is fundamental when calculating the EOS with finite-A methods.

\subsection{Solution of DFT in a periodic box}
\label{sec: skyrme box}

We discuss in detail the solution of the DFT problem for a finite number of nucleons enclosed in a cubic box with PBCs. 
We focus on spin-saturated PNM and SNM, which are the most important cases for nuclei and neutron stars \cite{rocamaza2018}. Moreover, SNM and PNM can be treated as two-component (spin up/down) fermionic systems in a unified way. 
The case of asymmetric matter ($\rho_n \ne \rho_p$, $\rm{N}\ne \rm{Z}$) would require some limited extensions of the formalism and is left for future studies. From now on, for the sake of simplicity in the notation the isospin labels ($q$ or $t$) are suppressed. 
%Some further details are reported in App. \ref{app: expr nuclear edf}.

We consider an external potential $v(z)$ that is a function of the $z$ coordinate only. Thus, translational invariance is broken in the $z$ direction, but still holds in the $xy$ plane. 
In order to respect PBCs, $v(z)$ must be periodic as well. Moreover, we adopt the spin- and isospin-independent sinusoidal potential
%assume it is independent of spin and isospin and in practical applications the sinusoidal potential
\begin{equation}
\label{eq: vz}
    v(z) = 2 v_q \cos \left( q z \right)
\end{equation}
 with $q$ being an integer multiple of $q_{min}=2\pi/L$.
The s.p. wave functions (in 2-spinor notation), then, have the following structure:
\begin{equation}
\label{eq: orbitals init}
    \psi_{\mathbf{n},\lambda}(\varX) = \frac{e^{i k_x x}}{\sqrt{L}} \frac{e^{i k_y y}}{\sqrt{L}} \, 
    \begin{pmatrix}
        \phi_{\mathbf{n},\lambda}(z,\uparrow) \\
        \phi_{\mathbf{n},\lambda}(z,\downarrow)
    \end{pmatrix}
\end{equation}
PBCs imply that $k_x$ and $k_y$ are quantized in units of $2\pi/L$, i.e. $k_x = \frac{2\pi}{L} n_x$ and $k_y = \frac{2\pi}{L} n_y$, and $\phi_{\mathbf{n},\lambda}(z)$ is periodic, i.e. $\phi_{\mathbf{n},\lambda}(z+L)=\phi_{\mathbf{n},\lambda}(z)$. 
The states are labelled by the three integer numbers $\mathbf{n}$, plus a spin quantum number $\lambda=\pm 1$ whose precise meaning will be discussed below. 

The general DFT equations \eqref{eq: skyrme HF eqs} are now specialized to our case.
We first note that the fields are functions of the z coordinate only: $m^{*}=m^{*}(z)$, $U=U(z)$ and $\mathbf{W}= W(z) \hat{\mathbf{z}}$. (The detailed expressions of the EDF and the fields are reported in App. \ref{app: expr nuclear edf}.)
For later convenience, we define the transverse momentum as
\begin{align}
\label{eq: def transverse momentum}
    \mathbf{k}_{n_x n_y} = k_x \hat{\mathbf{x}} + k_y \hat{\mathbf{y}} = \frac{2\pi}{L} \left( n_x \hat{\mathbf{x}} + n_y \hat{\mathbf{y}} \right)
\end{align}
having magnitude
\begin{align}
    k_{n_x n_y} = \sqrt{k_x^2 + k_y^2} = \frac{2\pi}{L} \sqrt{n_x^2 + n_y^2}.
\end{align}
Now, we discuss the spin-orbit term of Eq. \eqref{eq: skyrme HF eqs} with the help of $\pdv{\psi_{\mathbf{n},\lambda}}{x} = ik_x \psi_{\mathbf{n},\lambda}$ and $\pdv{\psi_{\mathbf{n},\lambda}}{y} = ik_y \psi_{\mathbf{n},\lambda}$:

\begin{align}
\label{eq: spin orbin intermediate}
    & \mathbf{W}(\varX) \cdot 
    \left( -i\right) \left( \nabla \cross \mathbf{\sigma} \right) \psi_{\mathbf{n},\lambda}(\varX) = \\
    & W(z) \left( -i\right) \left( \partial_x \sigma_y - \partial_y \sigma_x \right) \psi_{\mathbf{n},\lambda}(\varX) = \nonumber \\
    & W(z) \left( k_x \sigma_y - k_y \sigma_x \right) \psi_{\mathbf{n},\lambda}(\varX) = \nonumber \\
    & W(z) K_{n_x,n_y} \psi_{\mathbf{n},\lambda}(\varX) \nonumber
\end{align}
In the last equality, we have introduced the spin matrix $K_{n_x,n_y}=k_x \sigma_y - k_y \sigma_x $, which reads explicitly as
\begin{align}
    K_{n_x,n_y} =
    \begin{pmatrix}
    0               & -i (k_x + i k_y) \\
    i (k_x - i k_y) & 0
    \end{pmatrix} .
\end{align}
Since $K_{n_x,n_y}$ is not diagonal, it is clear that the states $\psi_{\mathbf{n},\lambda}$ cannot be eigenstates of $\sigma_z$. 
While one possibility would be to solve the coupled DFT equation for the spin-up and -down components, a better choice is to take the $\psi$'s to be eigenstates of $K_{n_x,n_y}$, as suggested in Ref. \cite{SemiInfMatter2008}. 
It is easy to verify that $K_{n_x,n_y}$ has eigenvalues $\pm k_{n_x n_y}$. 
Thus we impose
\begin{align}
\label{eq: helicity final}
    K_{n_x,n_y} \psi_{\mathbf{n},\lambda}(\varX) = \lambda k_{n_x n_y} \psi_{\mathbf{n},\lambda}(\varX), 
\end{align}
where $\lambda = \pm 1$.
Importantly, since $K_{n_x,n_y}$ is independent of the position, Eq. \eqref{eq: helicity final} implies that the orbitals \eqref{eq: orbitals init} can be decomposed into the product of a single spatial orbital and a constant spinor, namely 
\begin{equation}
\label{eq: orbitals}
    \psi_{\mathbf{n},\lambda}(\varX) = \frac{e^{i k_x x}}{\sqrt{L}} \frac{e^{i k_y y}}{\sqrt{L}} \, \phi_{\mathbf{n},\lambda}(z) \, \chi_{n_x, n_y, \lambda}.
\end{equation}
The spinors $\chi_{n_x,n_y,\lambda}$ satisfy
\begin{equation}
\label{eq: helicity eigenstate}
    K_{n_x,n_y} \chi_{n_x,n_y,\lambda} = \lambda k_{n_x n_y} \chi_{n_x,n_y,\lambda},
\end{equation}
where
\begin{equation}
    \chi_{n_x, n_y, \lambda} = \frac{1}{\sqrt{2}} 
\begin{pmatrix} 1 \\ \lambda e^{i\phi}
\end{pmatrix} \ .
\end{equation}
In the last expression, the angle $\phi$ is given by $\phi = \arctan\left( n_y / n_x \right)$.

Physically, the states $\psi_{\mathbf{n},\lambda}$ have a definite spin projection in the direction of the transverse momentum \eqref{eq: def transverse momentum}, which is not fixed but depends on the numbers $n_x$, $n_y$. The label $\lambda$ thus can be interpreted as a spin projection or helicity quantum number.
%\FM{In passing, it is interesting to mention a connection to the Dirac equation of interested in two-dimensional condensed matter models \cite{DiracSolidstate} and to the concept of helicity eigenstates.}

The kinetic term can be manipulated along the same lines and is discussed in App. \ref{app: kinetic term}. Finally, applying Eqs. \eqref{eq: kin term}, \eqref{eq: spin orbin intermediate} and \eqref{eq: helicity final} to Eq. \eqref{eq: skyrme HF eqs}, we find the following one-dimensional equations for the spatial orbital $\phi_{\mathbf{n},\lambda}(z)$:   
\begin{align}
\label{eq: skyrme final eqs}
    & -  \frac{d}{dz} \left(
   \frac{\hbar^2}{2m^{*}(z)} \phi_{\mathbf{n},\lambda}'(z)
   \right)  + \\ &
   \left( U(z) + v(z) + \lambda k_{n_x n_y} W(z) + \frac{\hbar^2}{2m^{*}(z)} k_{n_x n_y}^2 
   \right) \phi_{\mathbf{n},\lambda}(z) = \nonumber  \\
   &  \epsilon_{\mathbf{n},\lambda}
   \phi_{\mathbf{n},\lambda}(z) \nonumber .
\end{align}
These are s.p. state-dependent \Sch equations that must be solved self-consistently due the density-dependence of the fields. For a given set of quantum numbers $n_x$,$n_y$ and $\lambda$, $n_z$ labels the eigensolutions ordered by increasing s.p. energies $\epsilon$. The $z$ coordinate is restricted to the symmetric interval $\left[ -\frac{L}{2}, \frac{L}{2} \right]$.

We note that due to time-reversal invariance, that holds if we consider 
the spin-independent potential \eqref{eq: vz}, the eigenvalues $\epsilon_{\mathbf{n},+1}$ and $\epsilon_{\mathbf{n},-1}$ are degenerate, while in general $\lambda=\pm 1$ spatial orbitals are different.  
In the special case of homogeneous matter ($v=0$ and $\rho(z)=\rho_0$), though, the spin-orbit field $W(z)$ vanishes [see Eq. \eqref{eq: spin orbit field}], and thus the equations for the spin-orbit partners $\lambda=\pm 1$ are identical and so are the orbitals, namely $\phi_{\mathbf{n},+1}=\phi_{\mathbf{n},-1}$. As a consequence, the spin-orbit density vanishes too [Eq. \eqref{eq: spin orbit density final}] and thus uniform matter is insensitive to spin-orbit.
In passing, we also observe that the energy of a spin-saturated and closed-shell system is invariant when the sign of the spin-orbit coefficient is flipped, $C^{\nabla J} \longrightarrow - C^{\nabla J}$. Indeed, the effect of this transformation is that of
swapping the $\lambda=1$ and $\lambda=-1$ states in Eq. \eqref{eq: skyrme final eqs} and, if an equal number of spin states is occupied, all the densities, including $J(z)$, remain unchanged, and so does the total energy.

%We note that in the case of homogeneous matter ($v=0$ and $\rho(z)=\rho_0$) the spin-orbit field $W(z)$ vanishes (see Eq. \eqref{eq: spin orbit field}) and thus the equations for $\lambda=\pm 1$ are identical and so are the spatial orbitals, namely $\phi_{\mathbf{n},+1}=\phi_{\mathbf{n},-1}$. As a consequence, the spin-orbit density vanishes too [Eq. \eqref{eq: spin orbit density final}], as it should \cite{Marino2021}. 

We shall describe how the \Sch equation \eqref{eq: skyrme final eqs} is solved, how the many-particle g.s. of the system is constructed, and how the s.c. loop is dealt with. 
Due to the intrinsic periodicity of the systems under study, expanding Eq. \eqref{eq: skyrme final eqs} in the plane waves basis (see e.g. Refs. \cite{IzaacComputationalQM,Martin2020}) allows to solve the problem very efficiently. Few tens of plane waves are typically enough to find converged results even for moderately strong perturbations; by contrast, the finite-difference approach used in Ref. \cite{Gezerlis2022Skyrme} requires a mesh of several hundreds points at least and a much more time-consuming diagonalization.
The orbitals are Fourier-expanded as $\phi(z)= \frac{1}{\sqrt{L}} \sum_k c_k e^{ikz}$ where again $k=\frac{2\pi}{L} n$ and the \Sch equation is recast into matrix form, namely 
\begin{align}
\label{eq: eig plane waves}
    \sum_{k'} \left( \tilde{h}_{\mathbf{n},\lambda} \right)_{k,k'} c_{k'} = \epsilon_{\mathbf{n},\lambda} c_k, 
\end{align}
where $\left( \tilde{h}_{\mathbf{n},\lambda} \right)_{k,k'}$ is the Hamiltonian matrix in the plane waves basis and is derived in App. \ref{app: hamiltonian in plane waves}.

\begin{comment}
is compactly written as the sum of the Fourier transforms of the kinetic and potential terms, namely
\begin{equation}
\label{eq: hamiltonian tilde}
    \tilde{h}_{k,k'} = k' \tilde{B}(k-k') k + \tilde{U}(k-k').
\end{equation}
with $\tilde{B}(k-k')$ and $\tilde{U}(k-k')$ defined as
\begin{align}
    & \tilde{B}(k-k') = \frac{1}{L} \int_{-L/2}^{L/2} dz\, e^{-i(k-k')z} \frac{\hbar^2}{2m^{*}(z)} , \\
    & \tilde{U}(k-k') = \frac{1}{L} \int_{-L/2}^{L/2} dz \, e^{-i(k-k')z} \\ &\left( U(z) + v(z) + \lambda k_{n_x n_y} W(z) + \frac{\hbar^2}{2m^{*}(z)} k_{n_x n_y}^2 \right) . \nonumber
\end{align}
\end{comment}

Nuclear DFT is based on an independent-particle picture and the many-particle g.s. configuration is found by occupying the first $\rm{A}$ energy levels of the system.
In order to determine them, Eqs. \eqref{eq: skyrme final eqs} are solved for several different combinations $(n_x,n_y)$, and separately for the two spin states $\lambda$ \cite{Gezerlis2022Skyrme}. 
Then, the solutions are collated and the lowest-energy states are filled up with $\rm{A}/2$ spin-up and $\rm{A}/2$ spin-down particles. (The discussion is limited to spin-saturated system.)
Energy levels are degenerate, since $n_x$ and $n_y$ only enter Eq. \eqref{eq: skyrme final eqs} in the combination $k_{n_x n_y} \propto n_x^2+n_y^2$, so that inverting the sign of $n_x$, $n_y$ or both, or exchanging the two numbers, leaves the equation invariant.
Such degeneracy $g_{n_x,n_y}$ can be exploited to reduce the computational load of the method, since we can restrict ourselves to the pairs $(n_x,n_y)$ with $0 \le n_x \le n_y \le n_{max}$.
It is good practice to choose at first a large value for $n_{max}$, though the following argument, which generalizes that of Ref. \cite{Gezerlis2022Skyrme}, allows to stop the search over the $(n_x,n_y)$ pairs sooner. 
Indeed, we observe that $k_{n_x n_y}$ enters Eq. \eqref{eq: skyrme final eqs} in the combination $\lambda k_{n_x n_y} W(z) + \frac{\hbar^2}{2m^{*}(z)} k_{n_x n_y}^2 $.
This contribution is positive when $k_{n_x n_y}$ satisfies the inequality
\begin{align}
\label{eq: kxy condition}
    k_{n_x n_y} > \Bar{k}_{n_x n_y} = \max_z \left( - \lambda \frac{ 2m^{*}(z)  W(z) }{\hbar^2}
    \right).
\end{align}
Then, provided that $k_{n_x n_y} > \Bar{k}_{n_x n_y}$, the lowest eigenvalue of Eqs. \eqref{eq: skyrme final eqs} increases as $k_{n_x n_y}$ increases.
Now, while one is iterating over the combinations $(n_x,n_y)$ (which must have been sorted according to increasing values of $n_x^2+n_y^2$), and separately for $\lambda=+1$ and -1, one  checks whether the lowest eigenvalue $\epsilon_{n_x,n_y,0,\lambda}$ is greater than the energy of the first $\rm{A}/2$ lowest-energy states found so far. In that case, the cycle can be stopped, since we are guaranteed by Eq. \eqref{eq: kxy condition} that the many-nucleon g.s does not receive contributions from higher $n_x^2+n_y^2$. 

Once the occupied orbitals and the corresponding s.p. energies have been found, the total energy and the densities (App. \ref{app: dens orbitals}) of the system are computed. 
\begin{comment}
Number density, kinetic density and spin-orbit density may be computed from their definitions as functions of the occupied orbitals \cite{Schunck2019} applied to the wave functions \eqref{eq: orbitals}. Eqs. \eqref{eq: nabla psi} and \eqref{eq: laplacian psi} are also used to find
\begin{align}
    \rho(z) &= \sum_j \abs{\psi_j(\varX)}^2 = \, \frac{1}{L^2} \sum_{\mathbf{n},\lambda} \abs{\phi_{\mathbf{n},\lambda}}^2(z) \\
    \tau(z) &= \sum_j \abs{\nabla \psi_j (\varX)}^2 \\ &= \frac{1}{L^2} \sum_{\mathbf{n},\lambda} \left( 
    \abs{\phi_{\mathbf{n},\lambda}'}^{2} + k_{n_x n_y}^2 \abs{\phi_{\mathbf{n},\lambda}}^2
    \right)   \nonumber \\
    J_z(z) &= \sum_j \psi_j^*(\varX)  \left( -i\right) \left( \nabla \cross \mathbf{\sigma} \right)_3 \psi_j(\varX) \\
    &= \sum_{\mathbf{n},\lambda} \psi_{\mathbf{n},\lambda}^*(\varX) K \psi_{\mathbf{n},\lambda}(\varX) \nonumber \\
    &= \frac{1}{L^2} \sum_{\mathbf{n},\lambda} \lambda k_{n_x n_y} \abs{\phi_{\mathbf{n},\lambda}(z)}^2 \nonumber
\end{align}
where only the z component of $\mathbf{J}$ matters and Eq. \eqref{eq: helicity eigenstate} has been used. 
\end{comment}
The total energy is evaluated in two ways, i.e. as an integral of the energy density,
\begin{equation}
\label{eq: e integral}
    E = L^2 \int_{-L/2}^{L/2} dz \mathcal{E}(z),
\end{equation}
and by means of
\begin{equation}
\label{eq: energy erea}
    E = \frac{1}{2} \left( T + \sum_j \epsilon_j \right) + E_{rea}.
\end{equation}
The rearrangement energy $E_{rea}$ and the energy density $\mathcal{E}(z)$ are given in App. \ref{app: expr nuclear edf}. The expressions \eqref{eq: e integral} and \eqref{eq: energy erea} must match when they are evaluated on the g.s. and this provides a strong check on the correctness of the method and on its convergence to the exact g.s.

A crucial aspect of DFT is that the potential is itself a functional of the densities. Therefore, a s.c. solution to the problem must be looked for \cite{Schunck2019}. 
At each iteration $i$ of the s.c. loop, the densities are determined for the current values of the fields, as described above. 
Then new fields are generated by linearly mixing the old fields with the ones evaluated on the newly obtained densities $\rho^{(i)}$ \cite{SkyrmeRpa}, namely
\begin{align}
    U^{(i+i)} = \alpha U^{(i)}  + \left( 1-\alpha \right) U\left[ \rho^{(i)} \right]
\end{align}
and similar relations for $W$ and $\hbar/(2m^*)$. $\alpha$ is a mixing parameter; in order to achieve convergence, it is safe to be rather conservative, e.g. we choose $\alpha=0.8-0.9$ at the beginning and then gradually decrease it as iterations go by.
At the beginning ($i=0$), the densities are initialized at the uniform matter values $\rho(z)=\rho_0$, $\tau(z) = \frac{3}{5} \rho_0 q_F^2$ and $J(z)=0$ and the fields are determined accordingly.

The s.c. procedure is stopped if two conditions are met: the energies between iterations $i$ and $i-1$ and, at the same time, the two formulas \eqref{eq: energy erea} and \eqref{eq: e integral} for the energy at iteration $i$, agree within a chosen tolerance. Thresholds of the order of 0.1-1 keV per nucleon can be obtained usually in few tens of iterations. 
Combining linear mixing and two convergence conditions makes our approach rather robust.

\section{Theory of the static response}
\label{sec: static response theory}
The theory of the response of homogeneous matter to an external static perturbation is summarized. In-depth discussions can be found in Refs. \cite{giuliani_vignale_2005,SenatoreBook,Lundqvist}. 

Consider a system with uniform g.s. density $\rho_0$, described either by a Hamiltonian $\hat{H}$ or an EDF. A static potential $v(\varX)$ coupled to the total density is then turned on. $v(\varX)$ is periodic so as to respect the PBCs.
The density and energy of the g.s. of the perturbed system are called $\rho_v(\varX)$ and $E[v]$, respectively. 
If the external potential is weak enough, its effect can be treated perturbatively (see e.g. Refs. \cite{Fetter,giuliani_vignale_2005}). The density fluctuation induced by $v(\varX)$, in particular, is linear in the external potential and is written as follows:
\begin{align}
\label{eq: delta rho real space}
    \delta\rho(\varX) = \rho_v(\varX) - \rho_0 = \int d\varX' \chi(\varX,\varX') v(\varX').
\end{align}
The static response function $\chi(\varX,\varX')$ has been introduced and we stress that it depends exclusively on the properties of the unperturbed system. The response of homogeneous matter, in particular, is a function only of $\varX-\varX'$, i.e $\chi(\varX,\varX')=\chi(\varX-\varX')$.

While a generic periodic function $v(\varX)$ is a superposition of plane waves, in the following we consider without loss of generality a monochromatic potential oscillating at a given wave number $\varQ$, namely 
\begin{align}
\label{eq: periodic v}
    v(\varX) = v_q e^{i\varQ \cdot \varX} + c.c. = 2 v_q \cos\left( \varQ \cdot \varX \right).
\end{align}
Thus the density fluctuation induced by the perturbation \eqref{eq: periodic v} is monochromatic too and is given by
\begin{align}
\label{eq: delta rho harmonic}
    \delta \rho (\varX) = 2 \rho_q \cos\left( \varQ \cdot \varX \right),
\end{align}
where the amplitude $\rho_q$ is linear in $v_q$, i.e. 
\begin{align}
\label{eq: rhoq v}
    \rho_q = \chi(q) v_q
\end{align}
and $\chi(q)$ is the Fourier transform of $\chi(\varX,\varX')$, see Eq. \eqref{eq: chi fourier transform}.
The energy of the perturbed system, instead, is quadratic in the external potential. In App. \ref{app: static response theory}, we derive that the energy per particle is given by \cite{SenatoreBook}
\begin{equation}
\label{eq: ev quadratic}
    \delta e_v = e_v - e_0 = \frac{\chi(q)}{\rho_0} v_q^2.    %+ C_4 v_q^4
\end{equation}

The formalism we have outlined is valid both in the TL and in finite systems, and both for DFT and for Hamiltonian-based methods. 
The question is now how to compute the response function in practice.
For generalized Skyrme EDFs \cite{PASTORE20151} and Gogny and Nakada EDFs \cite{Pastore2021}, for example, the response in the TL can be determined analytically %, see the works by Pastore \etal \cite{PASTORE20151,Pastore2021} 
(App. \ref{app: edf tl resp}). 
An alternative for studying $\chi(q)$ is provided by exploiting Eqs. \eqref{eq: rhoq v} or \eqref{eq: ev quadratic}.
The strategy to determine $\chi(q)$ for a uniform system at a given density $\rho_0$, and with a given particle number, is the following.
For a given (quantized) momentum $q$,  multiple calculations of the g.s. of the perturbed system are performed for different values of the strength $v_q$ of the external potential \eqref{eq: periodic v}. Then $\chi(q)$ can be extracted from the amplitude of the density fluctuations [Eq. \eqref{eq: rhoq v}] or from  the energies  [Eq. \eqref{eq: ev quadratic}] as a function of $v_q$, for sufficiently small $v_q$. 
This strategy has been applied in several contexts, e.g. Refs. \cite{SenatoreBook,Gezerlis2017,Dornheim2017,Stringari2019}, and provides a relatively straightforward way to determine the static response function numerically.
We will interpolate energies using the more general formula \cite{Gezerlis2017,Dornheim2017}
\begin{align}
\label{eq: fourth order ev}
    \delta e_v = e_v - e_0 = \frac{\chi(q)}{\rho_0} v_q^2 + C_4 v_q^4
\end{align}
which takes into account higher-order contributions.

Second-order perturbation theory, or equivalently the spectral representation of the dynamical density response $\chi(\varQ,\omega)$, can be employed to derive a formula that relates $\chi(q)$ to the excited states of the homogeneous system \cite{Fetter,giuliani_vignale_2005}. 
For the case of the spin- and isospin-saturated $\rm{A}$-fermion FG, the response $\chi_{0,\rm{A}}$ at zero temperature is given by \cite{Dornheim2017,giuliani_vignale_2005}
\begin{comment}
\begin{equation}
    \chi_{0,A}(q) = \frac{1}{\Omega} \sum_{\varK,\sigma} \frac{n_{\varK+\varQ,\sigma}-n_{\varK,\sigma} }{\epsilon_{\varK+\varQ}-\epsilon_{\varK}}
\end{equation}
with $\sigma$ labels the spin projection and $n_{\varK,\sigma}$ and $\epsilon_{\varK}=\frac{\hbar^2\varK^2}{2m}$ are the occupation factors and the s.p. energies of the momentum eigenstates, respectively. 
At zero temperature and for spin/isospin saturated systems, the previous formula leads to
\end{comment}
\begin{equation}
\label{eq: chi0N analytic}
    \chi_{0,\rm{A}}(q) = - \frac{4mg}{\hbar^2 \Omega} \sum_{\varK\,\rm{occ}} \frac{1}{\left( \varK+\varQ\right)^2 - \varK^2 },
\end{equation}
where the sum extends over the occupied momentum states and terms with vanishing denominator are can be safely neglected.
Consistently with the assumptions of Sec. \ref{sec: dft formalism}, we write $\varK = \frac{2\pi}{L} \mathbf{n}$ and take $\varQ$ quantized and parallel to the $z$ direction, i.e. $\varQ=q \hat{\mathbf{z}} = \frac{2\pi}{L} p \, \hat{\mathbf{z}}$, with $p$ integer. 
Then Eq. \eqref{eq: chi0N analytic} is expressed as 
\begin{align}
\label{eq: chi0N analytic final}
    \chi_{0,\rm{A}}(q) = - \frac{mg}{L \pi^2 \hbar^2} \sum_{\mathbf{n}\,\rm{occ}}\frac{1}{p^2 + 2 p n_z}.
\end{align}
This formula is straightforward to evaluate: we determine the occupied states of the $\rm{A}$-particle FG g.s. once and then, for each value of $q$, we simply perform a sum over these states.
In the TL, $n_\varK=\theta(q_F-k)$, $\frac{1}{\Omega} \sum_\varK \longrightarrow \int \frac{d\varK}{(2\pi)^3}$ \cite{Fetter} and the static response becomes the well-known Lindhard function at zero-frequency \cite{Lindhard}
\begin{align}
\label{eq: lindhard tl}
    & \chi_0(q) = - g \frac{mq_F}{2 (\hbar \pi)^2} f\left( \frac{q}{2q_F} \right) \\
    \label{eq: fk lind}
    & f(k) = \frac{1}{2} \left( 1 + \frac{1-k^2}{2k} \log \abs { \frac{1+k}{1-k } } \right).
\end{align}

\section{Results}
\label{sec: results}
The method described in Sec. \ref{sec: dft formalism} is applied to calculate the EOS and the static response.
The popular SLy4 EDF \cite{Chabanat_Sly} is used when not stated otherwise, and examples of perturbed matter calculations are typically performed at a reference density of $\rho_0$= 0.16 fm$^{-3}$. DFT energies are converged within a tolerance of 1 keV per nucleon. 
Perturbation strengths are measured in units of the Fermi energy of the corresponding system ($v_q/E_F$).
We plot the static response function in the form $-\chi(q)/\rho_0$ (in MeV$^{-1}$), which is everywhere positive.
%while the static response function is normalized by the density and  changed of sign so to be positive everywhere ($-\chi(q)/\rho_0$, in MeV$^{-1}$).
Momenta are reported either in units of the Fermi momentum ($q/q_F$) or as integer multiples of the minimum allowed momenta ($q_{min}=2\pi/L$).

\subsection{EOS}
\label{sec: results eos}
\begin{figure*}[t]
    \begin{minipage}[t]{\columnwidth}
        \includegraphics[width=\columnwidth]{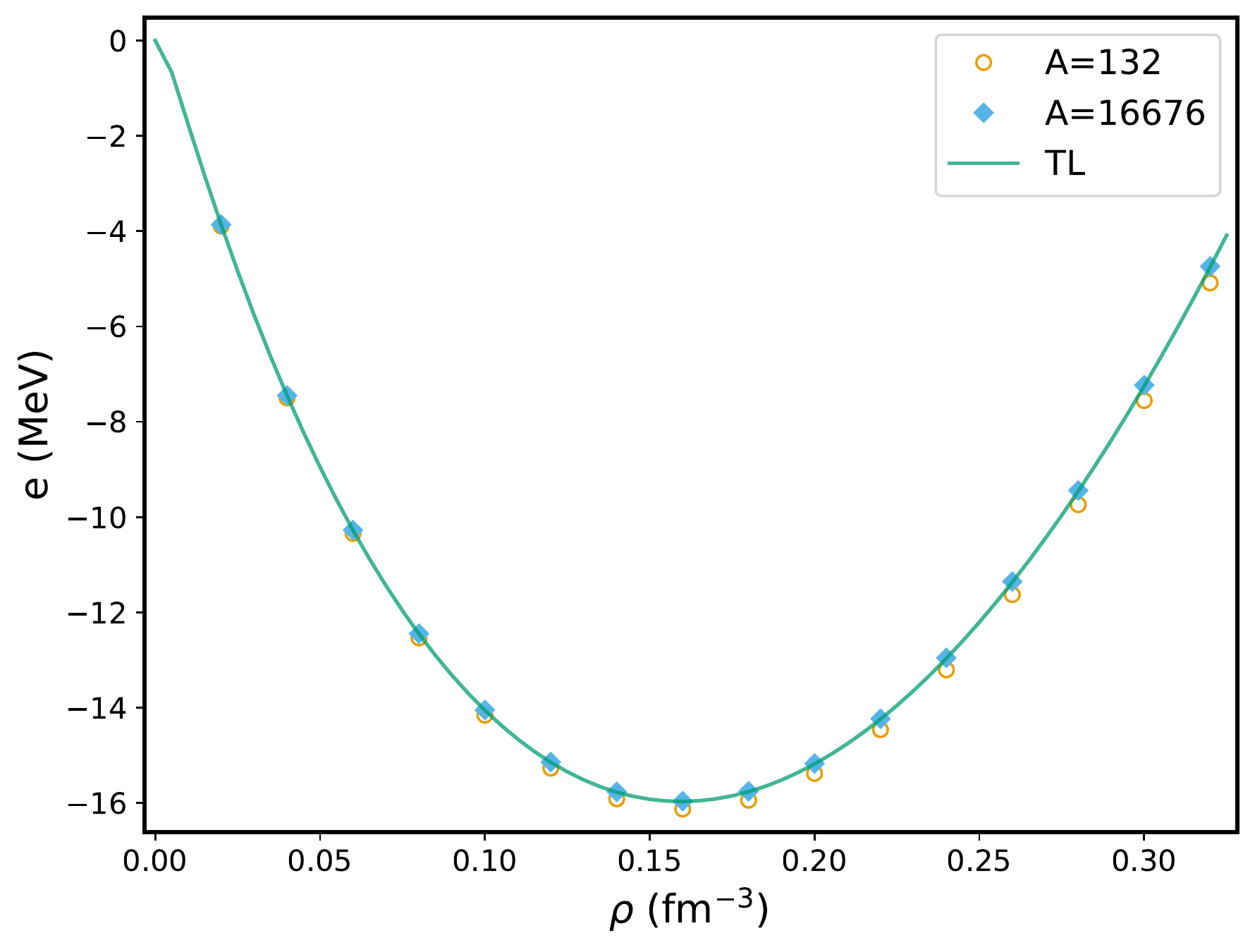}
    \caption{SNM EOS computed with the SLy4 EDF in the TL (line) and with a finite number of particles (symbols). }
    \label{fig: eos_sly4_snm}
    \end{minipage}
    \begin{minipage}[t]{\columnwidth}
    \includegraphics[width=\columnwidth]{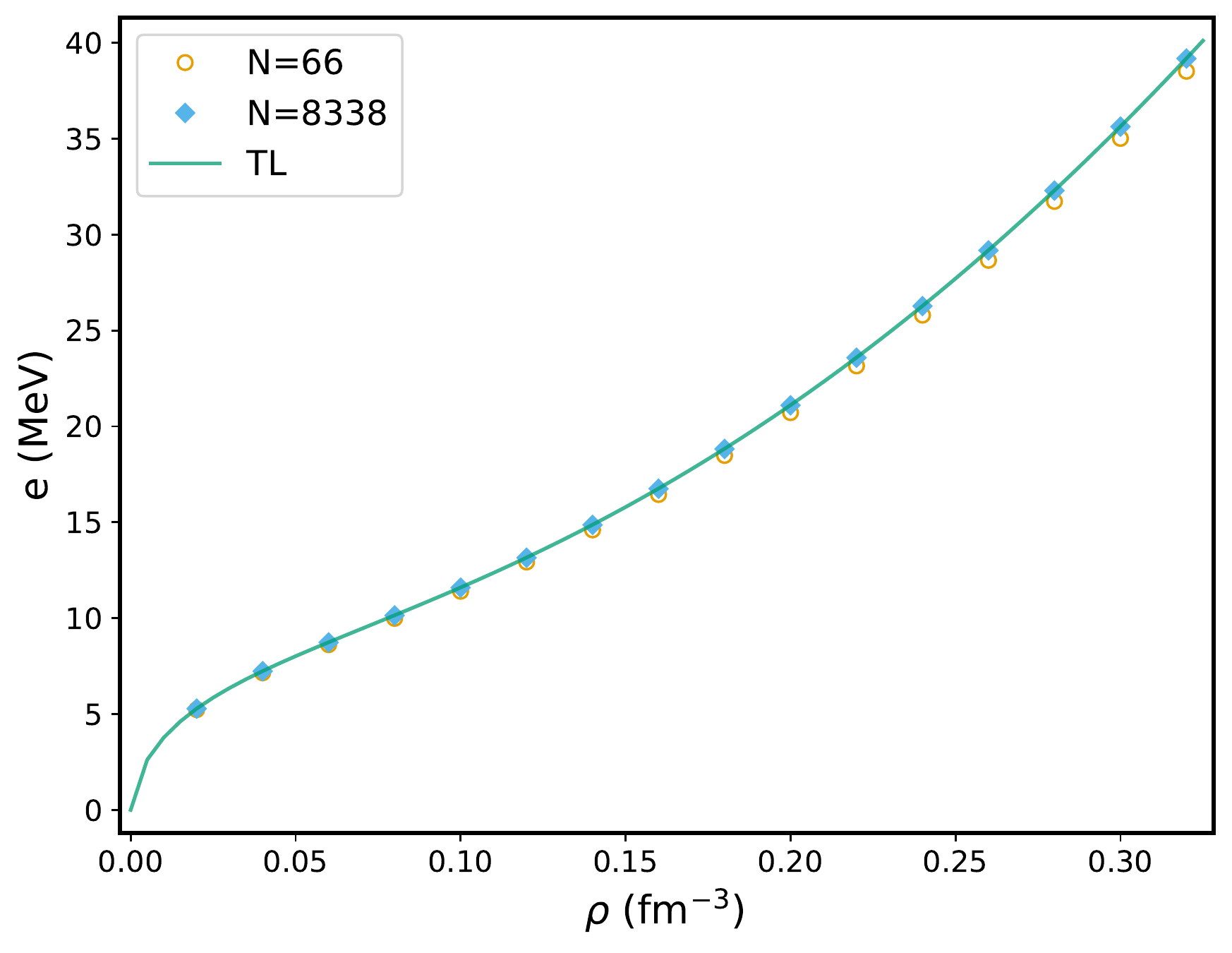}
    \caption{Same as Fig. \ref{fig: eos_sly4_snm}, but for PNM.
    }
    \label{fig: eos_sly4_pnm}
    \end{minipage}
\end{figure*}

As a first application, the EOS is studied in both SNM (Fig. \ref{fig: eos_sly4_snm}) and PNM (Fig. \ref{fig: eos_sly4_pnm}). The TL EOS is shown as a solid line, while calculations with $\rm{A}$=132, 16676 nucleons and $\rm{N}$=66, 8338 neutrons, respectively, are reported as symbols. Multiples of 33 particles are commonly used in infinite matter studies, because the kinetic energy per particle of FG made of 33$g$ particles is rather close to TL FG energy (see Ref. \cite{Gezerlis2017}, Fig. 1). 
As a prototypical large-$\rm{A}$ system, we use a number of nucleons equal to 4169 times the spin/isospin degeneracy $g$, which corresponds to filling up all the momentum shells of the FG up to $n^2=n_x^2+n_y^2+n_z^2=100$.
%By  filling up all the momentum shells of the FG up to $n^2=n_x^2+n_y^2+n_z^2=100$ we find the magic number 4169 (times the spin/isospin degeneracy $g$), which we use throughout this work as a prototypical large-$\rm{A}$ system. 
Indeed, the results for these numbers of nucleons turn out to be practically indistinguishable from the TL curve and provide a strong check on the correctness of numerical calculations. 
It can also be appreciated that the $\rm{N}$=66 and $\rm{A}$=132 EOS give energies very close to the TL EOS, so that the special usefulness of these "magic numbers" is confirmed also for DFT calculations.

% By contrast, the SNM EOS for $\rm{A}$=76, shown in Fig. \ref{fig: extrap_eos_sly4_snm} (filled symbols), clearly overbinds the TL EOS, with discrepancies that tend to increase as a function of the density. A simple formula for correcting the results of a finite-$\rm{A}$ method and extrapolate them towards the TL is the following \cite{Gezerlis2017}:
% \begin{align}   e_{corr} = e(A) - e_{FG}(A) + e_{FG}(TL), \end{align}
%where $e_{FG}$ denotes the free gas kinetic energies of $\rm{A}$ and infinite nucleons. The rationale is that the major contribution to FS effects should be due to the kinetic energy, that is strongly affected by the shell structure of finite systems, while potential terms should be less dependent on the particle number. The energies corrected by this formula are labelled as $\rm{A}=76\,\rm{corr}$ (hollow symbols) in the plot and one can notice that deviations from the TL shrink significantly about up to the saturation point, while are still quite large at high densities. By comparing with Figs. \ref{fig: eos_sly4_snm} and \ref{fig: eos_sly4_pnm}, we can state that $\rm{A}/g$=33 remains a better approximation to the TL EOS, whenever these calculations are feasible.

\begin{comment}
\begin{figure}
    \centering
    \includegraphics[width=0.45\textwidth]{extrap_eos_sly4_snm.pdf}
    \caption{ SNM EOS in the TL (line), computed with $\rm{A}$=76 nucleons (filled symbols) and extrapolated from $\rm{A}$=76 to the TL (hollow symbols). See text.  }
    \label{fig: extrap_eos_sly4_snm}
\end{figure}
\end{comment}

\subsection{Free response}
\label{sec: results free response}
\begin{figure*}[t]
    \begin{minipage}[t]{\columnwidth}
\includegraphics[width=\textwidth]{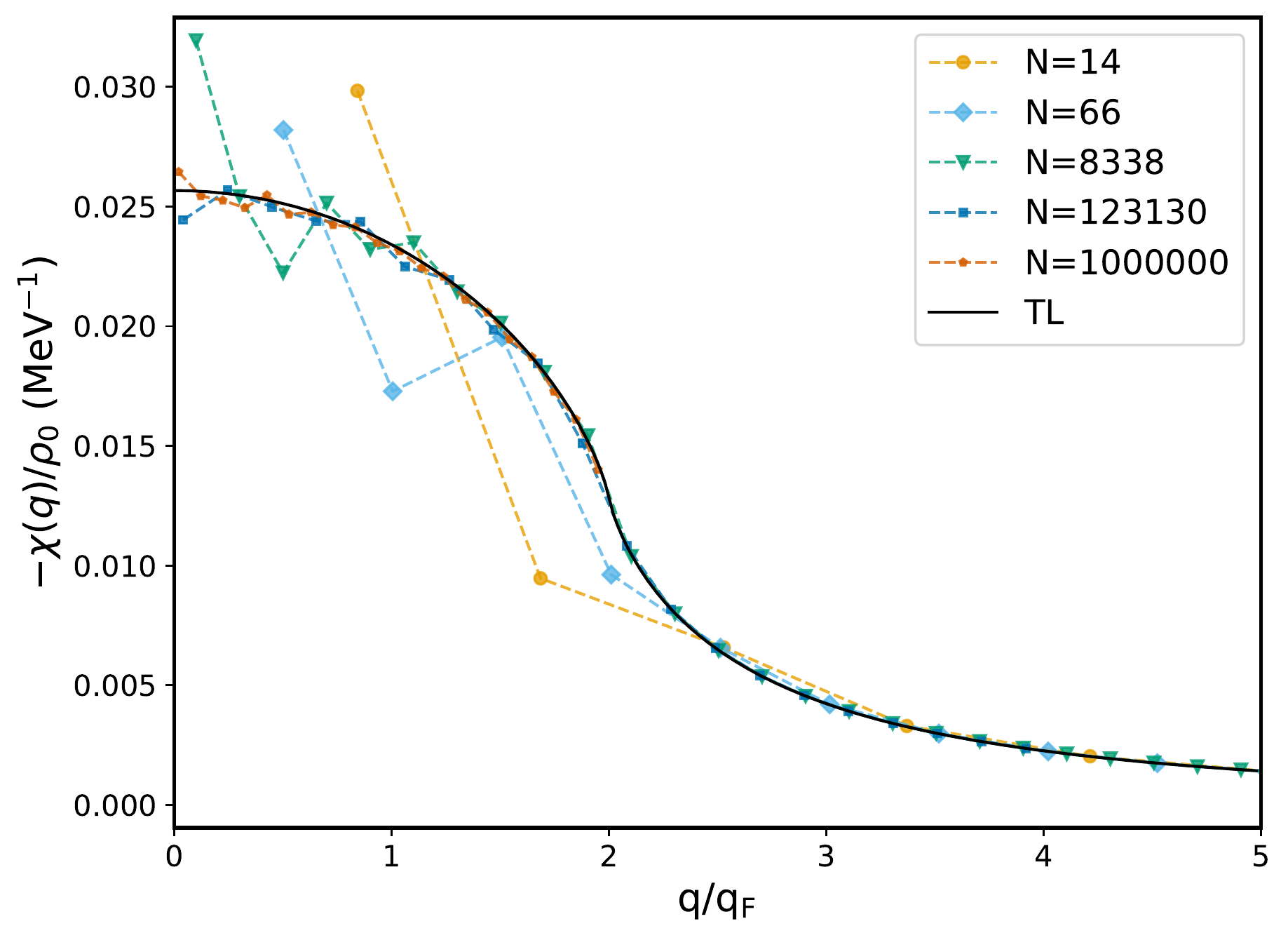}
    \caption{Dashed lines: free response function $-\chi_{0,N}(q)/\rho_0$ in PNM at $\rho_0=$0.16 fm$^{-3}$ as a function of $q/q_F$ for different numbers of neutrons. Full line: response in the TL (Lindhard function).  }
    \label{fig:free resp cinv}
    \end{minipage}
    \begin{minipage}[t]{\columnwidth}
    \includegraphics[width=\textwidth]{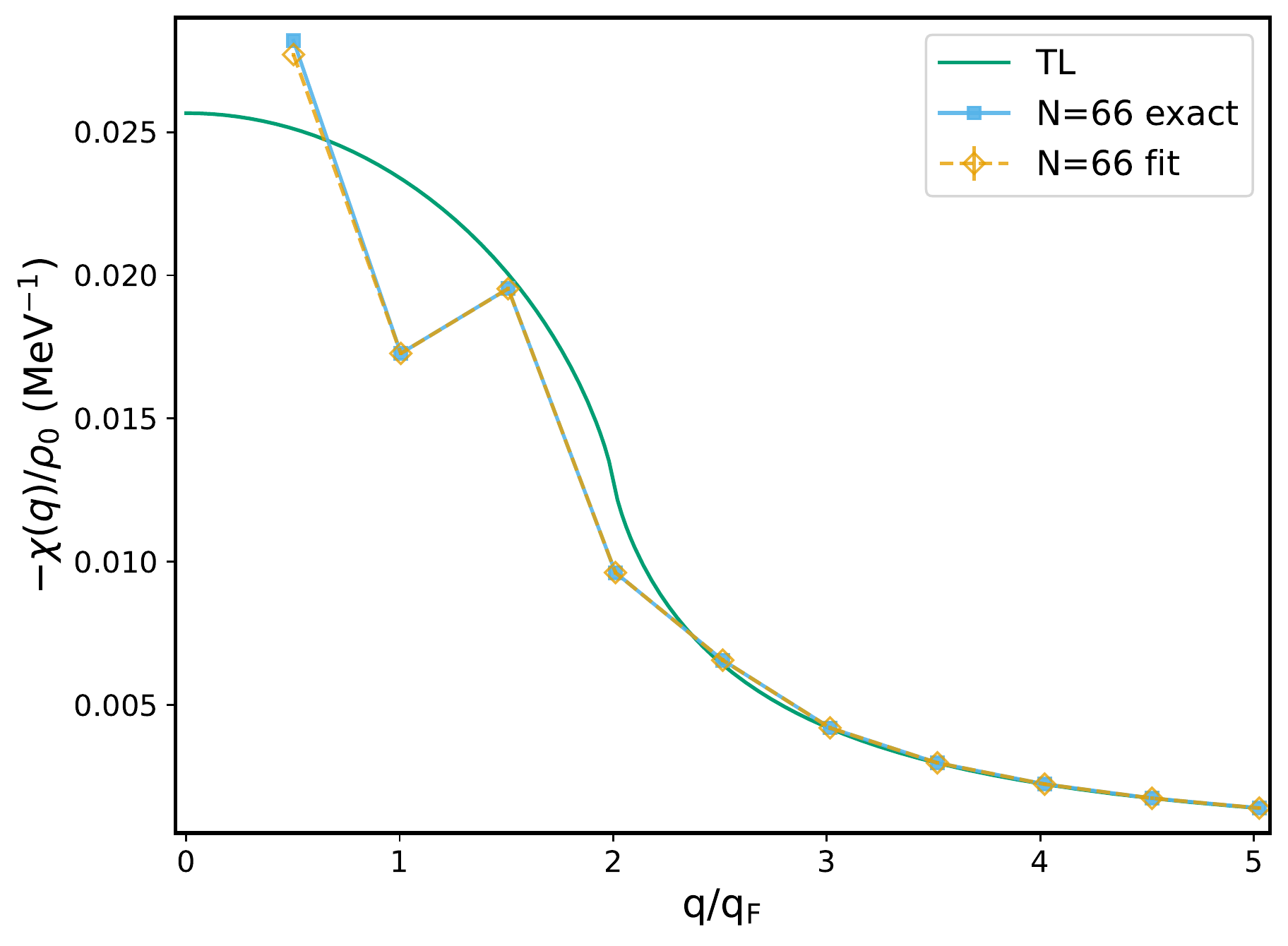}
    \caption{ Static response $-\chi_{0,\rm{N}}(q)/\rho_0$ of the FG as a function of $q/q_F$ in PNM at a density $\rho_0=0.16$ fm$^{-3}$.
    The exact response (filled squares) and the response determined by a fit to the Mathieu energies (empty diamonds) are shown for  $\rm{N}$=66 neutrons. For comparison, the TL response (Lindhard function) is also plotted. }
    \label{fig:resp mathieu N66}
    \end{minipage}
\end{figure*}

A second study concentrates on the static response of the FG.
%, for which numerical results can be compared to the exact formula for $\chi_{0,A}$ [Eq. \eqref{eq: chi0N analytic}].
The exact formula for $\chi_{0,\rm{N}}$ [Eq. \eqref{eq: chi0N analytic}] is applied in Fig. \ref{fig:free resp cinv} for different numbers of neutrons and compared to the TL response \eqref{eq: lindhard tl}. FS effects are rather strong at small or moderate momenta and manifest themselves as a non-monotonic behaviour of $\chi_{0,\rm{N}}(q)$ at finite $\rm{N}$, while the TL response function is strictly decreasing in magnitude. For $q>2q_F$, instead, the oscillations tend to disappear and the curves match rather well for all particle numbers. This qualitative change of behaviour is due to geometric reasons, see e.g. the calculation of $\chi_0(q)$ in Ref. \cite{Fetter}: essentially, for $q>2q_F$ any occupied momentum state can be scattered from the g.s. (the Fermi sphere) to an empty state and thus shell effects, that strongly affect the results at small $q$, are ineffective. The special role of $q=2q_F$ is also signalled by
the fact that the TL Lindhard function \eqref{eq: lindhard tl} is non-analytical at that point.
%\sout{ a singularity in the} \FM{the fact that at that point the } TL Lindhard function \eqref{eq: lindhard tl} \FM{is non-analytical}.
Moreover, we note that the convergence to the TL as $\rm{N}$ is increased is relatively slow and mild oscillations continue to persist up to very large $\rm{N}$.

\begin{figure*}[t]
    \begin{minipage}[t]{\columnwidth}
    \includegraphics[width=\textwidth]{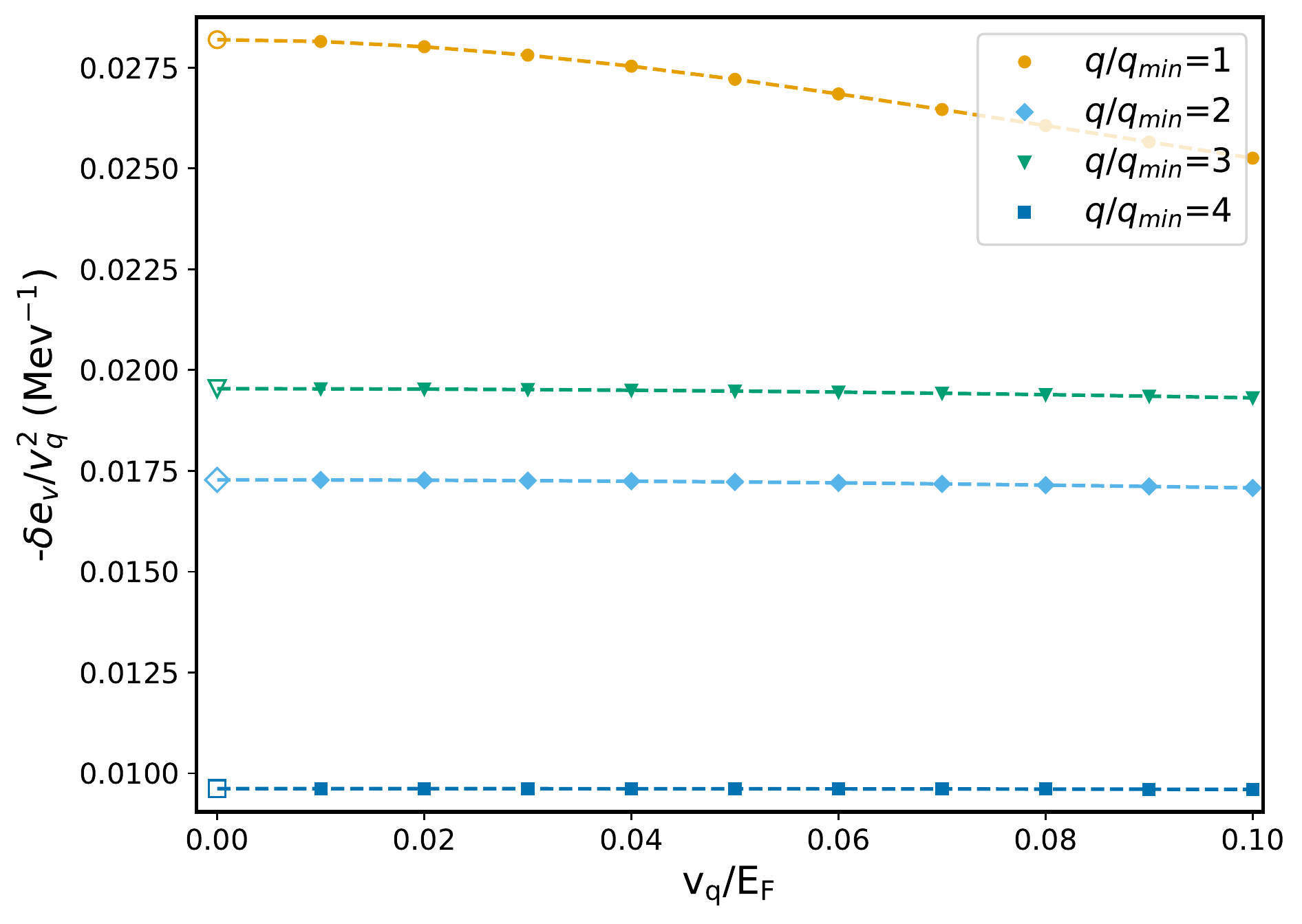}
    \caption{ Ratio between the energy variation $-\delta e_v$ and the square of the perturbation strength $v_q$ for the first four allowed moments ($q/q_{min}$ between 1 and 4) for the same system as Fig. \ref{fig:resp mathieu N66}.
    Hollow symbols at $v_q=0$ represent the exact value of -$\chi_{0,\rm{N}}(q)/\rho_0$.
    Dashed lines are guide to the eye.}
      \label{fig:ratio mathieu N66}
    \end{minipage}
    \begin{minipage}[t]{\columnwidth}
    \includegraphics[width=\textwidth]{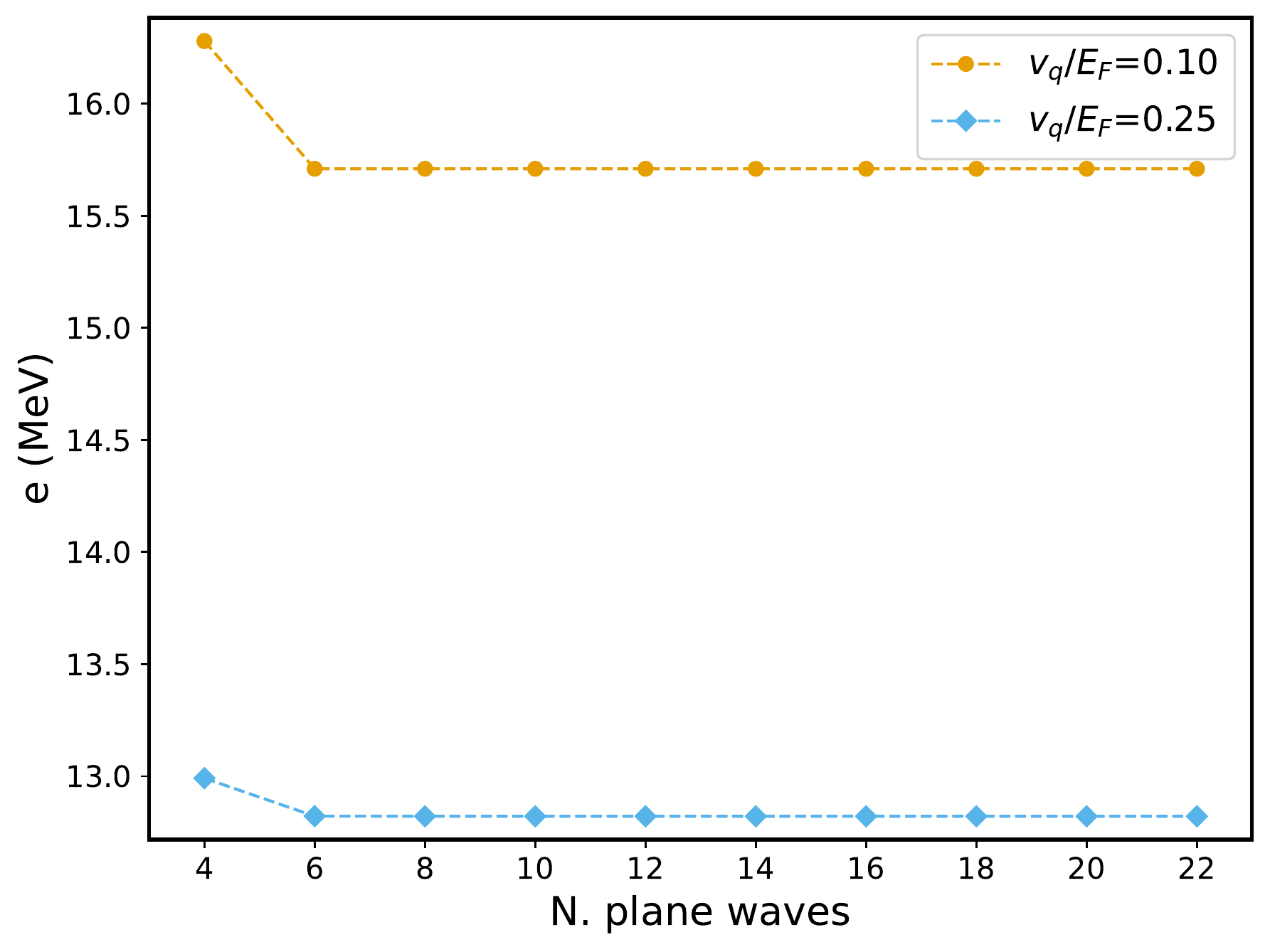}
    \caption{ Energy per particle of PNM with $\rm{N}$=66 at $\rho_0=0.16$ fm$^{-3}$ obtained with the SLy4 EDF as a function of the number of plane waves. Results are shown for the lowest momentum ($q=q_{min}$) for two different strengths of the external potential. }
    \label{fig: conv_nk_pnm_N66}
    \end{minipage}
\end{figure*}

Then, the free response is computed numerically and compared to the analytical results.
In particular, the FG response is determined by solving the Mathieu problem \cite{Gezerlis2017}, i.e. the independent-particle problem of fermions subject to the external potential \eqref{eq: vz} (with the EDF potential terms turned off), for different momenta $q$ and for strengths $v_q/E_F$ between 0.01 and 0.1 (with a step of 0.01). Then the energy differences $\delta e_v$ are interpolated with the quartic formula \eqref{eq: fourth order ev} at each $q$. 
% The method involves solving the Mathieu problem \cite{Gezerlis2017}, i.e. the EDF is turned off and the independent-particle problem of fermions subject to the potential $v(z)=2v_q \cos\left( q z \right)$ is solved. Then, for each periodicity $q$, the response function is found by fitting the energies as a function of the perturbation strength.In particular, energies $e_v$ for potentials $v_q/E_F$ between 0.01 and 0.1 (with a step of 0.01) are been calculated and interpolated. 
%With a pragmatic attitude, we try interpolating the energy differences $\delta e_v$ both with the quadratic formula \eqref{eq: ev quadratic} and with the fourth-order one \eqref{eq: fourth order ev} and present the estimate of $\chi(q)$ given by the best-performing fit (in terms of the standard reduced $\chi^2$) among the two. 
In Fig. \ref{fig:resp mathieu N66}, a comparison is drawn in the case of PNM with $\rm{N}$=66 neutrons between the exact response (filled squares) and the values obtained through the fitting procedure (empty diamonds). An almost perfect agreement is obtained, with a modest discrepancy only at the lowest momentum ($q/q_F \approx$ 0.5).
In order to better understand this deviation, in Fig. \ref{fig:ratio mathieu N66} we consider the ratio between the energy variation $\delta e_v$  and the square of the perturbation strength $v_q$ as a function of $v_q/E_F$. The exact response is shown as a hollow symbol at $v_q=0$.
If linear response theory were exact, at least in a certain range of small $v_q$, the ratio  $\delta e_v/v_q^2$ would be constant. This is indeed verified for $q/q_{min}>1$ over the whole interval considered, but at $q/q_{min}=1$
%while convergence of the ratio as $v_q \longrightarrow 0$ is correctly observed [$\delta v_q/v_q^2 \longrightarrow \chi_{0,\rm{N}}(q)/\rho_0$], 
a slight underestimation of the response is observed at all finite perturbations.
This highlights that modest non-linear (fourth-order) contributions are present in the behaviour of the system. Importantly, though, the ratio correctly converges to the exact response [  $\delta e_v/v_q^2 \longrightarrow \chi_{0,\rm{N}}(q)/\rho_0$] as $v_q \longrightarrow 0$. 

\subsection{Perturbed nuclear matter}
\label{sec: results perturbed matter}
\begin{figure*}
    \begin{minipage}[t]{\columnwidth}
        \includegraphics[width=\columnwidth]{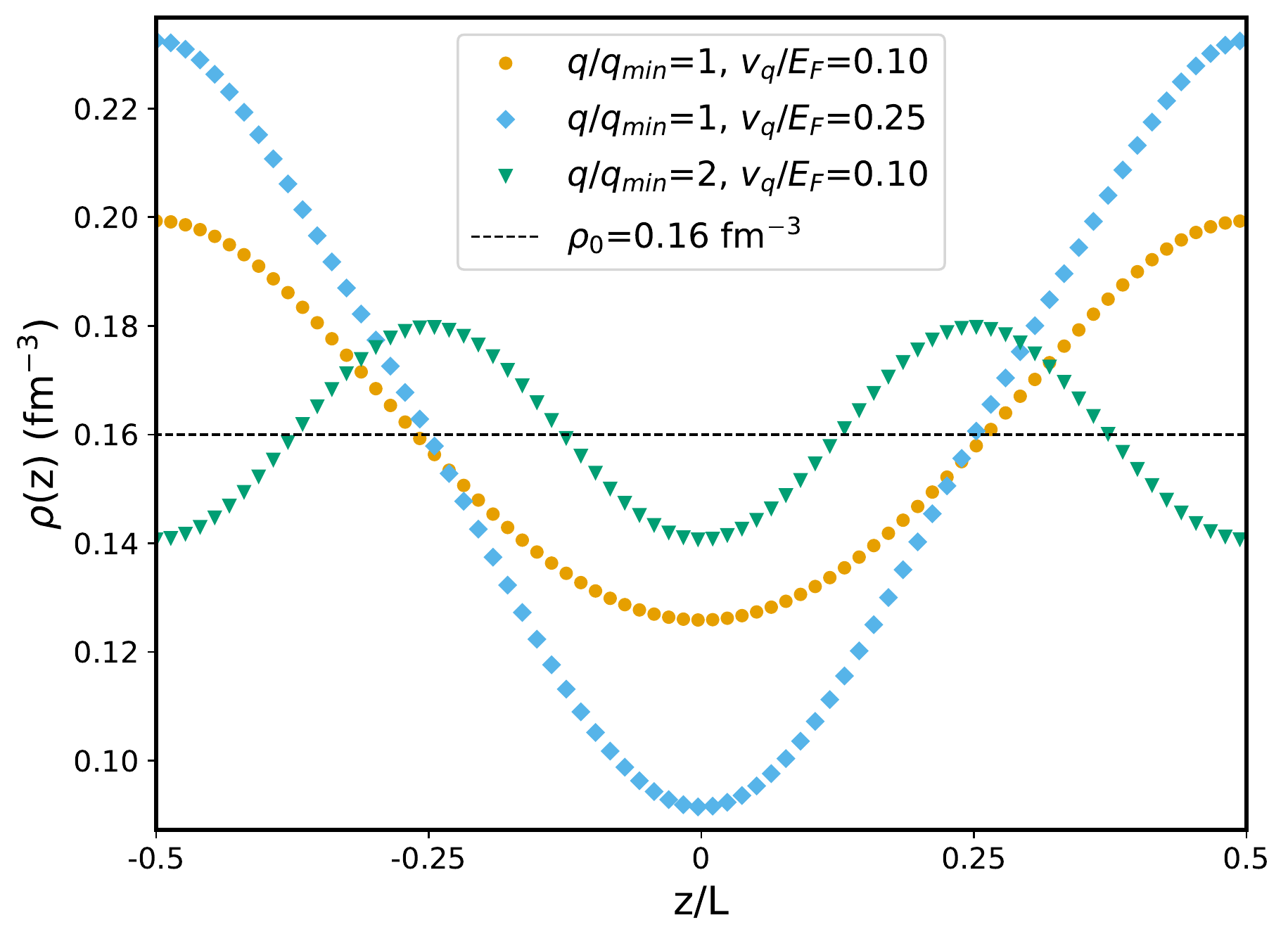}
    \caption{ Densities $\rho(z)$ as a function of $z/L$ in PNM ($\rm{N}$=66 neutrons) at a reference density $\rho_0=0.16$ fm$^{-3}$ (dashed horizontal line). Densities for three perturbations, differing in strength and momentum (see legend), are shown as symbols.  }
    \label{fig: densities_pnm_N66}
    \end{minipage}
    \begin{minipage}[t]{\columnwidth}
    \includegraphics[width=\columnwidth]{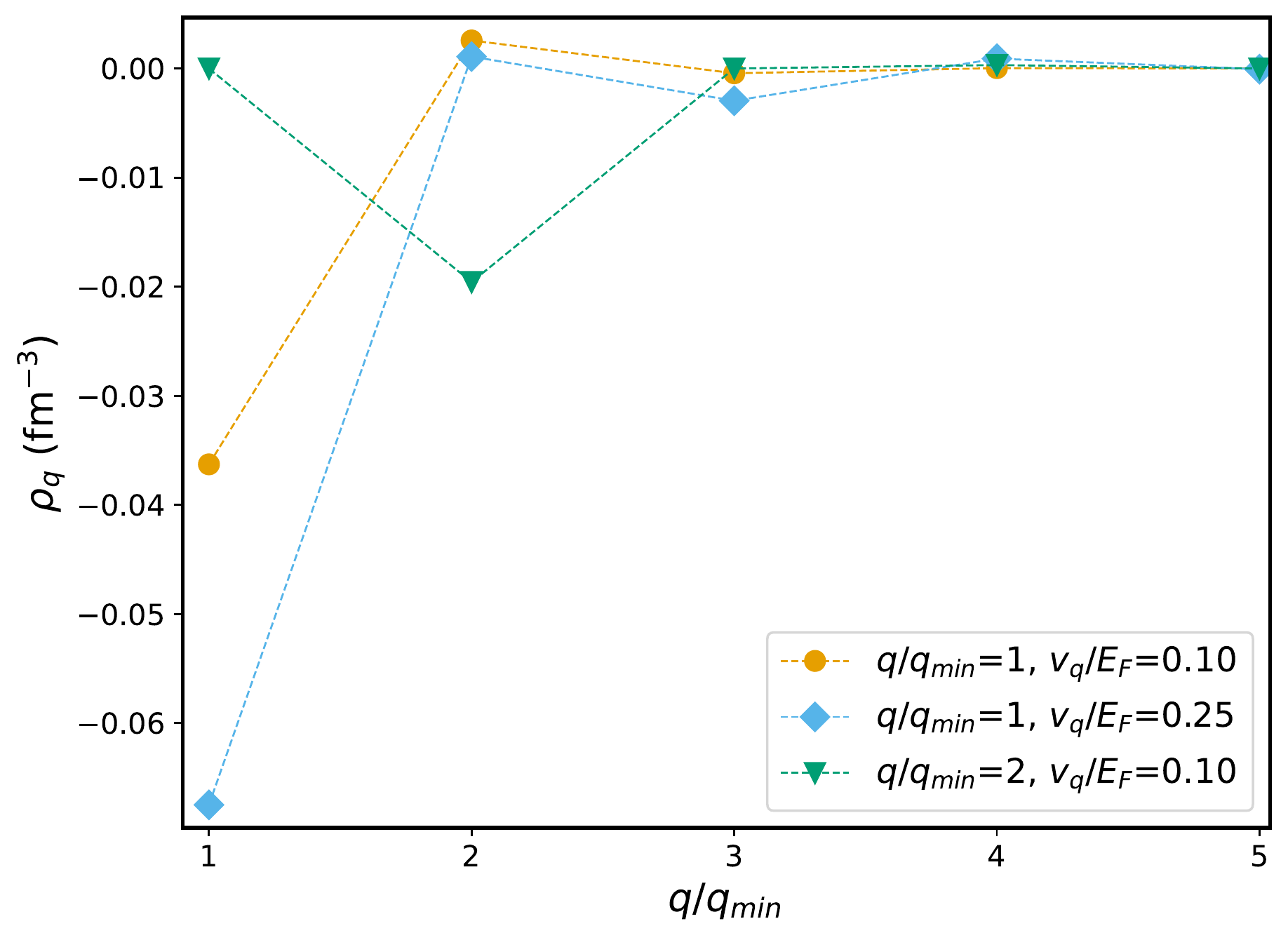}
    \caption{ Fourier components $\rho_q$ of the density fluctuations in the same cases as Fig. \ref{fig: densities_pnm_N66}.   }
    \label{fig: fourier_pnm_N66}
    \end{minipage}
\end{figure*}

Perturbed matter is now studied with the SLy4 EDF. 
First, a preliminary analysis of the convergence of the calculations with respect to the number of plane waves included in the basis is presented.  Fig. \ref{fig: conv_nk_pnm_N66}, which reports calculations performed with $\rm{N}$=66 neutrons at $q/q_{min}=1$ for a small ($v_q/E_F$=0.1) and a moderate ($v_q/E_F$=0.25) perturbation strengths, shows that in this case as few as 8 plane waves are sufficient to find energies converged within 0.1 keV or less. As a general rule, though, the number of plane waves required increases as a function of  the momentum $q$ of the perturbation and in practice we have found that a basis of 40 waves always yields converged results for 66 or 132 nucleons. When thousands of particles are considered, we raise the cutoff to 60 plane waves. Calculations remain very fast (few seconds) even on a single processor.
Then, the densities $\rho(z)$ as well as their Fourier components are shown in Figs. \ref{fig: densities_pnm_N66} and  \ref{fig: fourier_pnm_N66}, respectively, for three perturbations that differ in strength and periodicity ($q/q_{min}$=1 with strengths $v_q/E_F$=0.1, 0.25 and $q/q_{min}$=2 with $v_q/E_F$=0.1).
From the real space representation, one can appreciate that densities closely resemble cosine function that oscillate around the unperturbed density with the same periodicity as that of the external perturbation [see Eq. \eqref{eq: delta rho harmonic}]. The Fourier analysis confirms that the response is essentially harmonic, as in all cases a single component at momentum $q$ is clearly dominant with rather modest contributions beyond the linear regime.

\begin{figure}
    \centering
    \includegraphics[width=0.45\textwidth]{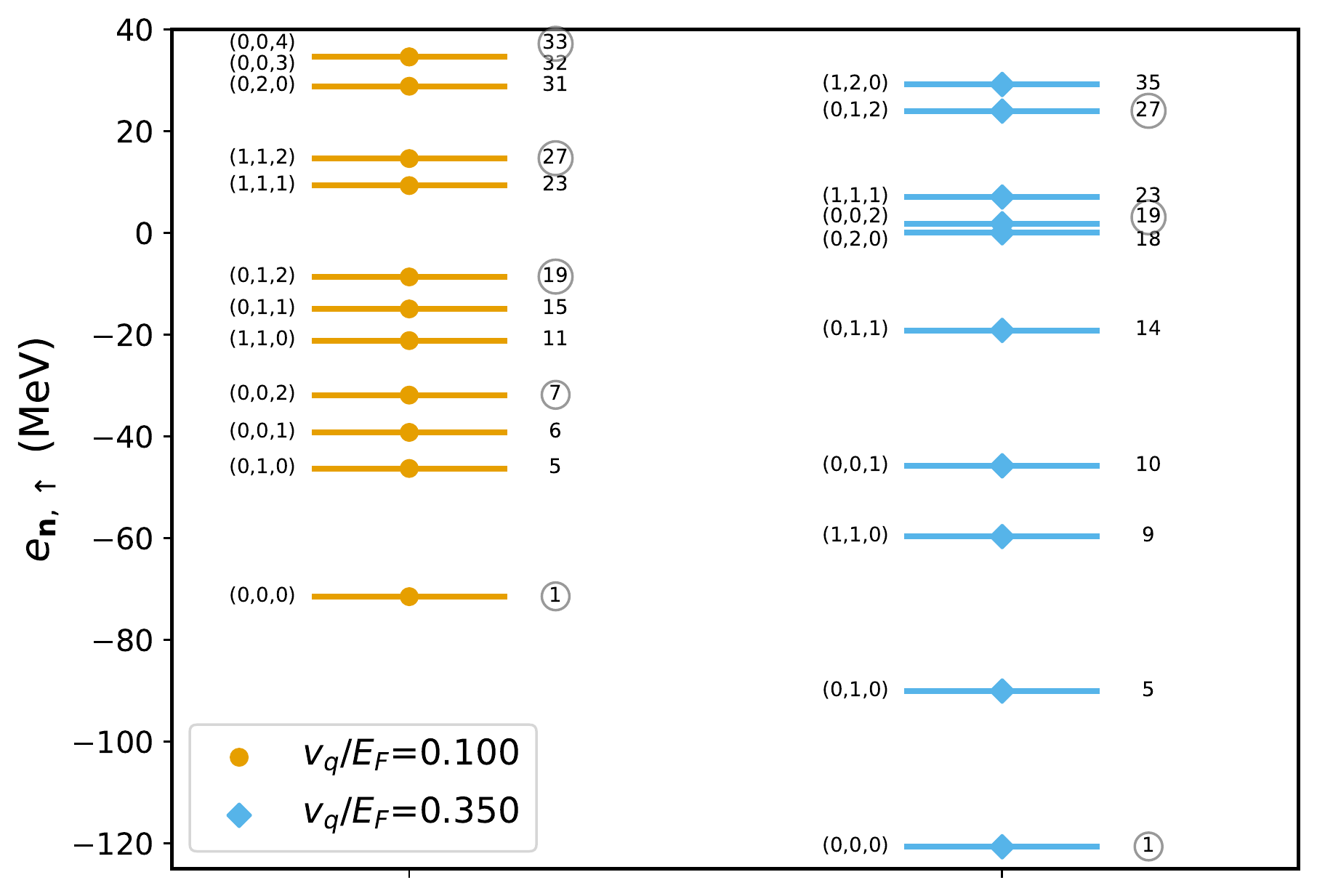}
    \caption{ Level structure of $\rm{N}$=66 PNM. Two perturbation strengths (at momentum $q/q_{min}=1$) are shown. The quantum numbers $\mathbf{n}=(n_x,n_y,n_z)$ of each level and the number of particles up to that shell are reported.
    Momentum-shell magic numbers of the FG are circled.  }
    \label{fig: levels}
\end{figure}

So far, we have always used particle numbers that correspond to a shell closure of the free Fermi gas and implicitly assumed that they are magic numbers for the perturbed system as well. This hypothesis proves true in general for weak potentials. Actually, its violation is a sign that the picture itself of a small perturbation of the homogeneous system is breaking down. 
In Fig. \ref{fig: levels} the neutron level scheme of $\rm{N}$=66 PNM (same case as Fig. \ref{fig: densities_pnm_N66}) is shown at two different perturbation strengths (both with momentum $q/q_{min}=1$). We remind that the $\lambda = \pm 1$ energy eigenvalues are degenerate and we plot the s.p. energies only for $\lambda=+1$.
The quantum numbers $\mathbf{n}=(n_x,n_y,n_z)$ ($0\le n_x \le n_y$), and the number of nucleons corresponding to shell closures, are reported next to each level. Among the latter, magic numbers of the FG are circled. 
In the case of the weaker potential, the effect of the perturbation is to partially lift the degeneracy of the free gas levels (as well as to lower the s.p. energies), as can be seen from the triplets or doublets of neighbouring levels. The overall structure of the homogeneous system, though, is preserved and indeed all the FG magic numbers up to 33 are found in the perturbed system too. 
A markedly different picture appears for the stronger perturbation, where the level ordering of the FG is severely altered. One consequence is that a shell closure is found not for 33 nucleons but for 35. We suggest that the sudden changes in the slope of the energy as a function of the perturbation mentioned in Ref. \cite{Gezerlis2022Skyrme} may be a side-effect of such 'shell-opening' effects.  
The key message is that care must be taken when studying perturbed finite-$\rm{A}$ matter and not only global properties (energy, density), but also the shell structure must be looked at. For example, we warn that, if DFT or Mathieu orbitals are used to construct a reference state for Quantum Monte Carlo \cite{Gezerlis2017,Lynn2019}, it is crucial to check that it be a closed-shell state, before embarking on expensive calculations.

\begin{figure*}[t]
    \begin{minipage}[t]{\columnwidth}
        \includegraphics[width=\columnwidth]{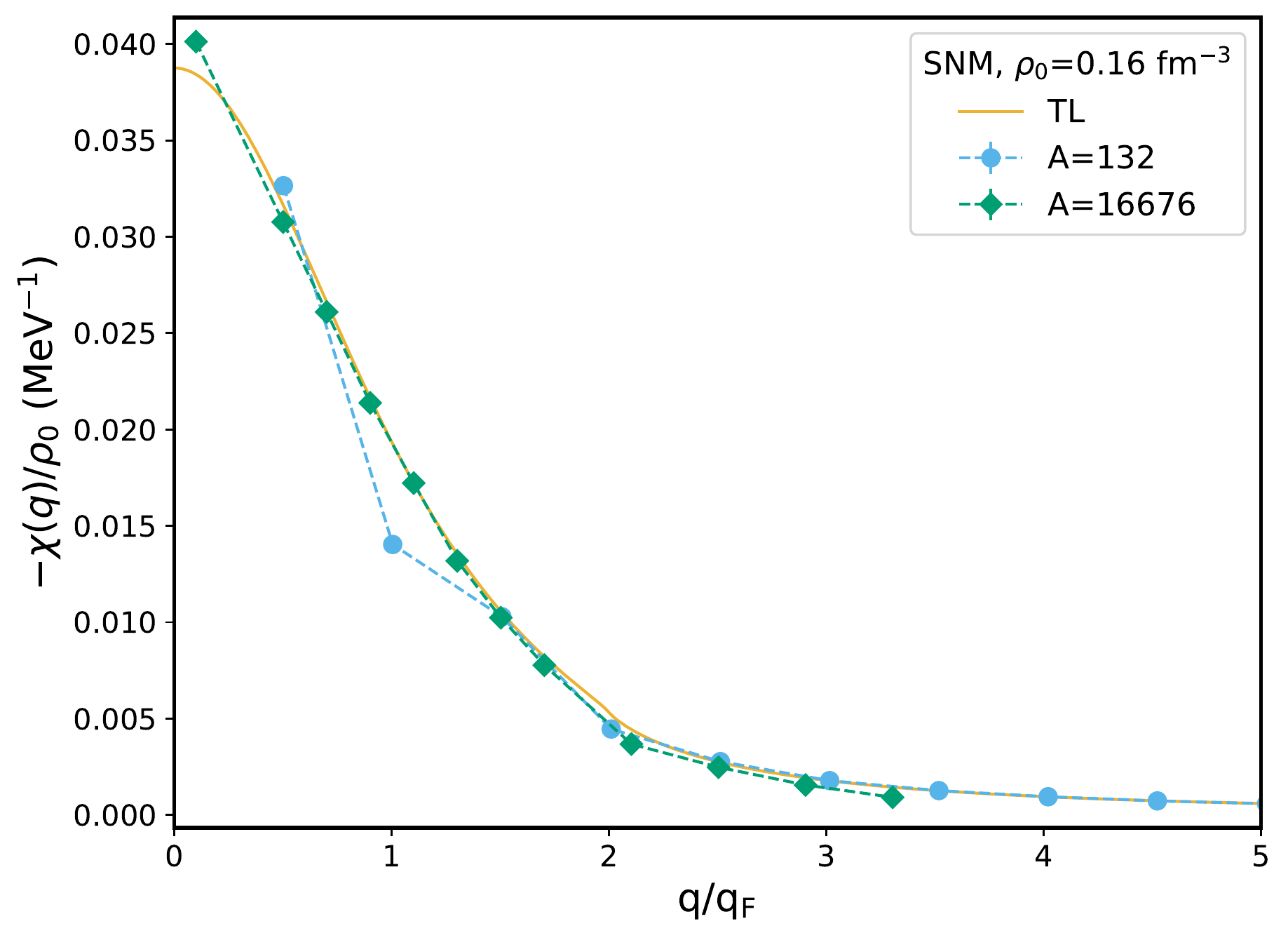}
    \caption{Static response of SNM at $\rho_0=0.16$ fm$^{-3}$ obtained with the SLy4 EDF. The solid line represents the TL response, while symbols denote calculations for a finite number of particles ($\rm{A}$=132 and 16676). }
    \label{fig:resp sly4 snm}
    \end{minipage}
    \begin{minipage}[t]{\columnwidth}
    \includegraphics[width=\columnwidth]{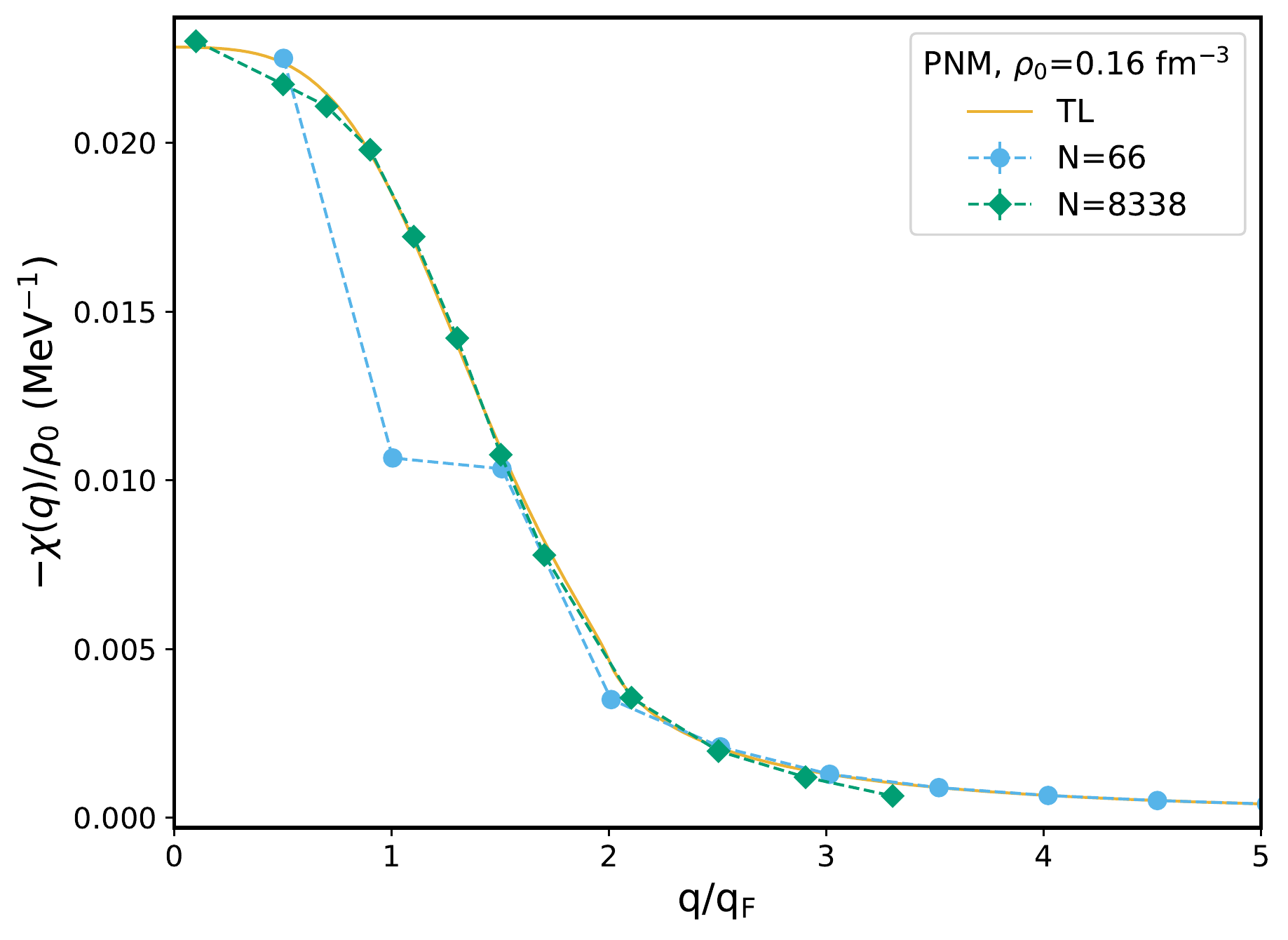}
     \caption{ Same as Fig. \ref{fig:resp sly4 snm}, but for PNM. Calculations are performed with $\rm{N}$=66 and 8338 neutrons (symbols) and in the TL. }
   \label{fig:resp sly4 pnm}
    \end{minipage}
\end{figure*}

%\subsection{DFT response}
%\label{sec: dft response}
%Perturbed DFT matter and the DFT response are discussed.
Next, the static response function is discussed. 
The TL response of nuclear EDFs is known exactly \cite{PASTORE20151} (App. \ref{app: edf tl resp}) and is now compared to the finite-$\rm{A}$ calculations in both SNM (Fig. \ref{fig:resp sly4 snm}) and PNM (Fig. \ref{fig:resp sly4 pnm}).
%Large-$\rm{A}$ calculations are performed in a basis of 60 plane waves and are rather fast (a few seconds) even on a single processor. 
The numerical response functions for the large-$\rm{A}$ system are in very good agreement with the analytical predictions.
The convergence to the TL is thus verified and we can appreciate by comparing to Fig. \ref{fig:free resp cinv} that it is definitely faster (as a function of the number of nucleons) in the interacting (DFT) system than for the FG.
The small-$\rm{A}$ response, instead, is characterized by a non-monotonic behaviour that is reminiscent of that of the free response, with marked fluctuations with respect to the TL function for $q<2 q_F$.

\begin{figure*}[t]
    \begin{minipage}[t]{\columnwidth}
        \includegraphics[width=\columnwidth]{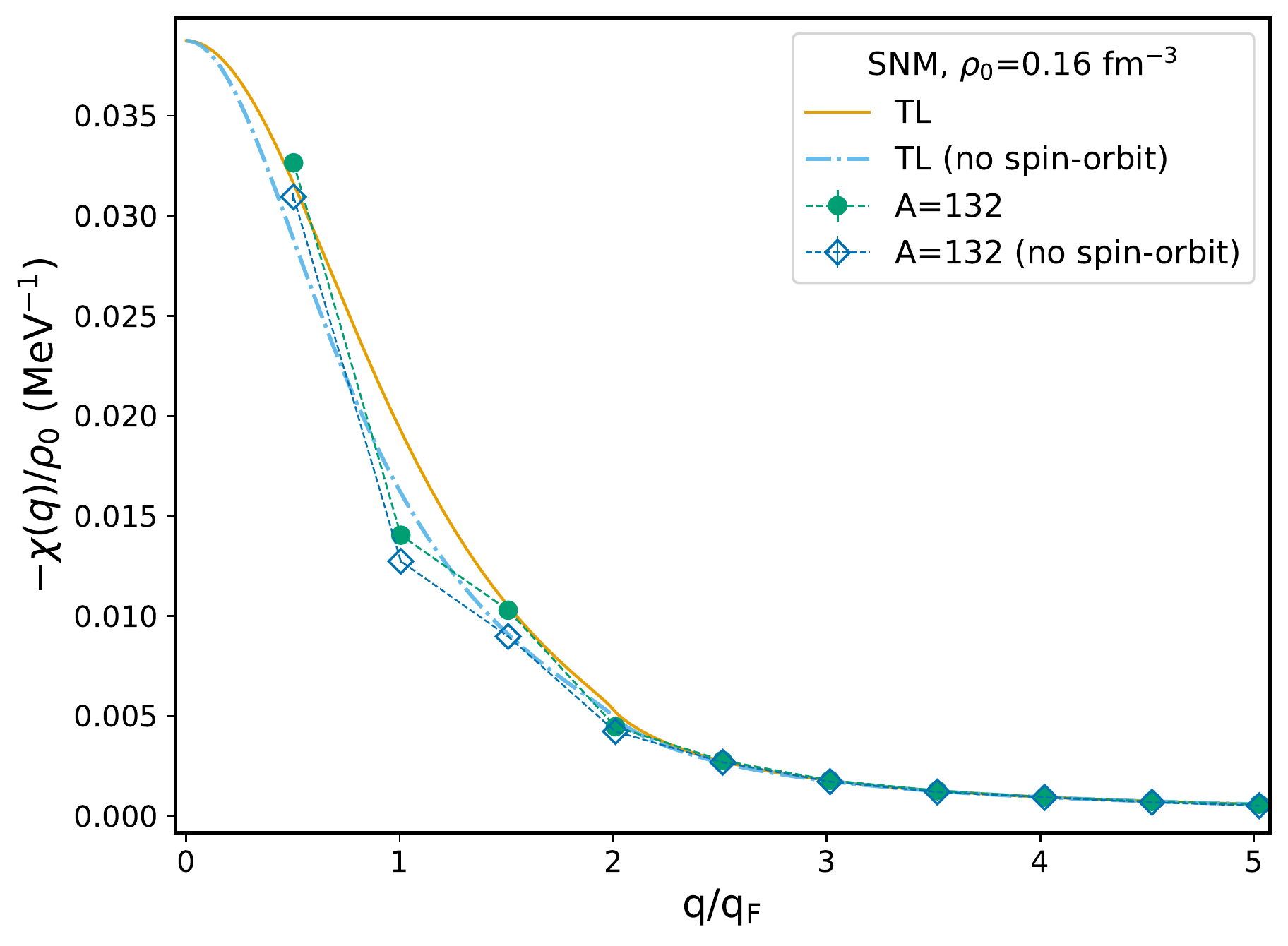}
    \caption{SNM static response obtained in the TL and for $\rm{A}$=132 nucleons with the full SLy4 EDF and SLy4 with spin-orbit terms neglected ('no spin-orbit' in the legend).   }
    \label{fig:spinorbit_sly4_snm}
    \end{minipage}
    \begin{minipage}[t]{\columnwidth}
    \includegraphics[width=\columnwidth]{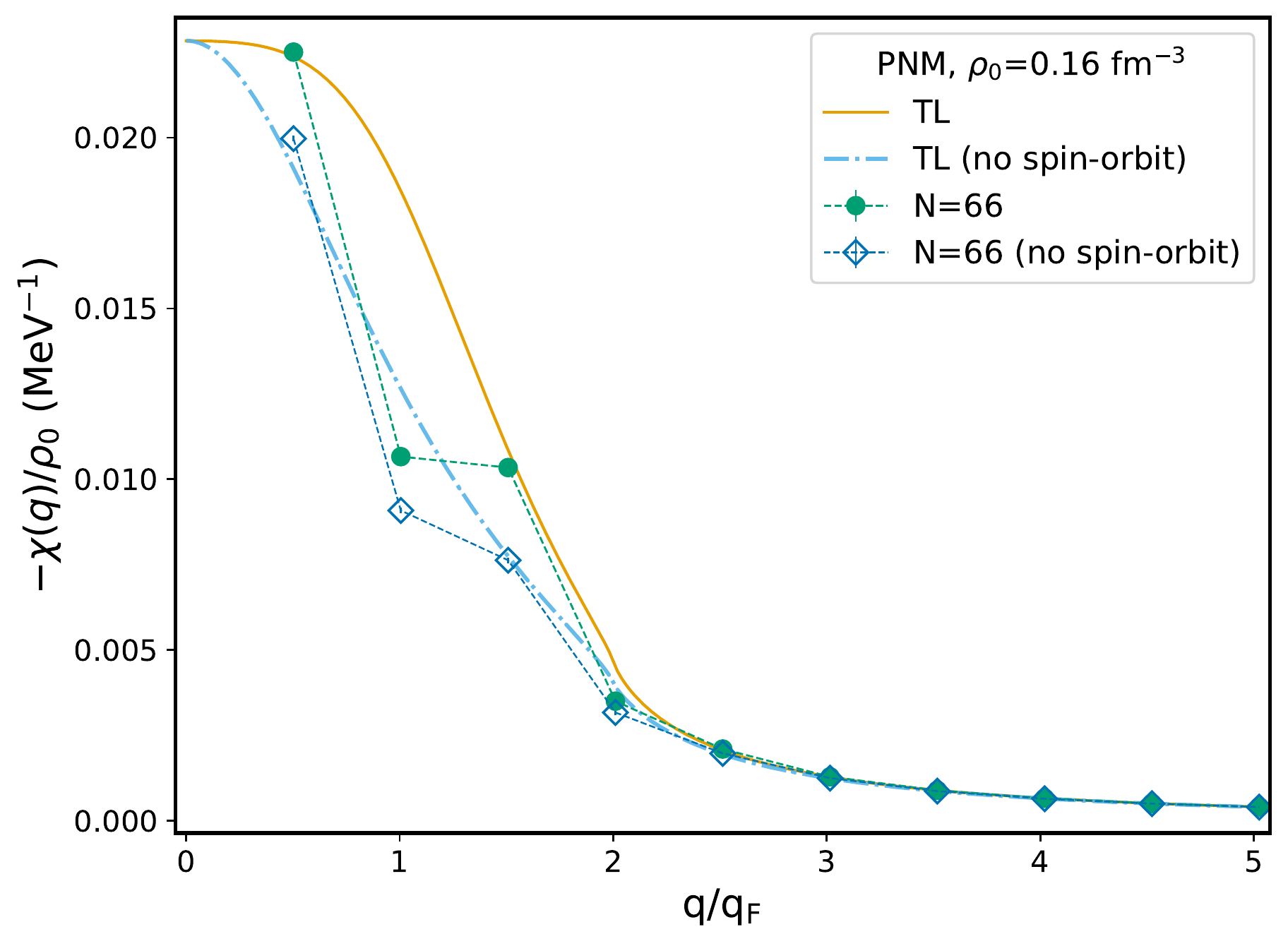}
     \caption{ Same as Fig. \ref{fig:spinorbit_sly4_snm}, but for PNM with $\rm{N}$=66 neutrons.  }
   \label{fig:spinorbit_sly4_pnm}
    \end{minipage}
\end{figure*}

Lastly, we would like to understand the impact of the spin-orbit terms on the static response. Spin-orbit was neglected in Ref. \cite{Gezerlis2022Skyrme} and its inclusion is one of the novelties of our work.
The response computed with the full SLy4 EDF and for SLy4 with spin-orbit neglected, i.e. with $C^{\nabla J}$ set to zero, is reported for SNM (Fig. \ref{fig:spinorbit_sly4_snm}) and PNM (Fig. \ref{fig:spinorbit_sly4_pnm}) both in the TL and for the usual $\rm{A}$=132 and $\rm{N}$=66 numbers of particles, respectively. 
One can appreciate that spin-orbit has the main effect of lowering the magnitude of $\chi(q)$ at all momenta, both in the TL and in the finite systems and, while in SNM it constitutes a small correction, in PNM it is a significant effect. While the qualitative picture of Ref. \cite{Gezerlis2022Skyrme} is not altered in a fundamental way, quantitative results may change noticeably. In particular, it is important to incorporate spin-orbit terms if one aims at constraining the EDF parameters using \textit{ab initio} information.

\section{Conclusions and perspectives}
\label{sec: conclusions}
To sum up, in this work we have studied nuclear matter under the effect of an external potential within the DFT framework. Our approach is based on simulating nuclear matter with a finite number of nucleons enclosed in a box and subject to PBCs, and the theoretical formalism and numerical implementation have been presented in detail for PNM and SNM for Skyrme-like EDFs.
We have discussed carefully how to treat spin-orbit terms and, in particular, we have shown that, although in the presence of spin-orbit the DFT orbitals are not eigenstates of the spin projection operator, single-component equations can still be derived.
Then, the problem of the response of nuclear matter to static density perturbations has been analyzed with our technique.

Our method has been validated successfully by comparing the numerical results with analytical formulas for the EDF EOS, the free gas response (both for finite-$\rm{A}$ and TL systems) and the TL EDF response. The power of DFT is demonstrated by the fact that systems of thousands of particles can be computed in an extremely fast and reliable way, and the convergence to the thermodynamic limit has been verified numerically.
Moreover, the validity of linear response for weak perturbations, as well as deviations occurring for stronger external potentials have been investigated by looking at energies, densities and level structures. We point out that the momentum space magic numbers of uniform matter do not necessarily correspond to shell closures of the perturbed system. Therefore, care must be taken when the finite-$\rm{A}$ DFT approach is used in conjunction with \textit{ab initio}, for example when DFT or Mathieu orbitals \cite{Gezerlis2017} are used as a reference state in Quantum Monte Carlo. 
Moreover, we have found that spin-orbit contributes significantly to the PNM response, and to a lesser extent to the SNM response. 
In future studies of inhomogeneous matter, therefore, spin-orbit terms should be incorporated.

% The method has been applied to the response of nuclear matter to static density perturbations. The static response has been studied extensively, judging the impact of finite-size effects, the convergence to the thermodynamic limit, and the effect of spin-orbit contributions.

This work represents an intermediate step in the program of developing \textit{ab initio}-based EDFs started in Ref. \cite{Marino2021}. 
Indeed, inhomogeneous systems are to be studied in order to gain information about the gradient terms of the EDF. Our efforts are currently devoted to the \textit{ab initio} response of both SNM and PNM, aiming at constraining the nuclear EDF by matching DFT and \textit{ab initio} results. In particular, our strategy involves tuning the EDF parameters on the \textit{ab initio} energies obtained with the same number of particles so to keep FS effects under control. Results will be presented in a forthcoming publication \cite{in_preparation}.

Moreover, while here we have focused on PNM and SNM and presented results for density perturbations only, the formalism can be easly extended to isospin-asymmetric matter, as well as (introducing time-odd densities in the theory \cite{Schunck2019}) to spin-polarized matter and to spin/isospin perturbations. 

\section{Acknowledgements}
We thank Alessandro Lovato and Francesco Pederiva for useful discussions.
F.M. acknowledges the use of CINECA Galileo100 computing resources through the AbINEF ISCRA-B grant.

\appendix
\section{Details on nuclear EDFs}
Further details on the EDF and the mean fields are provided. 
In this work we focus on PNM and SNM, that can be treated as two-component (spin up/down) fermionic systems. 
We adopt the convention for which $C^\tau$ stands for $C_0^\tau$ in SNM and $C_{nm}^\tau = C_0^\tau + C_1^\tau$ in PNM, and likewise for $C^{\Delta\rho}$, $C^{\nabla J}$ and the $c_\gamma$ coefficients. 

\subsection{EDFs}
\label{app: expr nuclear edf}
The expression of the EDF $\mathcal{E}$ under the assumptions of Sec. \ref{sec: dft formalism} is the following:
\begin{align}
    \mathcal{E}(z) &= \mathcal{E}_{kin}(z) + \mathcal{E}_{bulk}(z) +
    C^\tau \rho(z) \tau(z) + \\
    & C^{\Delta\rho} \rho(z) \rho''(z) 
     - C^{\nabla J} \rho'(z) J_z(z) \nonumber
\end{align}
with 
\begin{align}
    \mathcal{E}_{kin}(z) = \frac{\hbar^2}{2m} \tau(z),
    \\
     \mathcal{E}_{bulk}(z) = \sum_\gamma c_\gamma \rho^{\gamma+1}(z).
\end{align}
The rearrangement term was computed in Ref. \cite{Marino2021} and is given by
\begin{align}
    E_{rea} = L^2 \int dz \sum_\gamma 
    \left(
    \frac{1-\gamma}{2}
    \right) c_\gamma
    \rho^{\gamma+1}(z).
\end{align}
The expressions for the  mean field, effective mass and spin-orbit potential are also shown:
\begin{align}
     & \frac{\hbar^2}{2m^{*}(z)} = \frac{\hbar^2}{2m} + C^\tau \rho(z), \\
    & U(z) = U^{bulk}(z) + C^{\tau} \tau(z) + 2 C^{\Delta\rho} \rho''(z) + C^{\nabla J} J_z'(z)
\end{align}
with
\begin{align}
    U^{bulk} = \sum_\gamma c_\gamma \left( \gamma + 1 \right) \rho^\gamma(z),
\end{align}
and lastly
\begin{align}
\label{eq: spin orbit field}
    W_z(z) = - C^{\nabla J} \rho^{'}(z).
\end{align}

\subsection{Kinetic term} 
\label{app: kinetic term}
We derive the kinetic term of Eq. \eqref{eq: skyrme final eqs}. 
First, the gradient and the Laplacian of $\psi_{\mathbf{n},\lambda}$ [Eq. \ref{eq: orbitals}] are reported:
\begin{align}
\label{eq: nabla psi}
    \nabla \psi_{\mathbf{n},\lambda}(\varX) &= i k_x \psi_{\mathbf{n},\lambda}(\varX) \hat{\mathbf{x}} + i k_y \psi_{\mathbf{n},\lambda}(\varX) \hat{\mathbf{y}} \\
    & + \frac{1}{L} e^{i\left( k_x x + k_y y \right) } \chi_{n_x, n_y, \lambda} \phi_{\mathbf{n},\lambda}'(z) \hat{\mathbf{z}} \nonumber,
\end{align}
    
\begin{align}
\label{eq: laplacian psi}
    \nabla^2 \psi_{\mathbf{n},\lambda}(\varX) &= - \left( k_x^2+k_y^2 \right) \psi_{\mathbf{n},\lambda}(\varX) \\ &+ \frac{1}{L} e^{i\left( k_x x + k_y y \right) } \chi_{n_x, n_y, \lambda} \phi_{\mathbf{n},\lambda}''(z) \nonumber.
\end{align}
Using these expressions, we elaborate on $ - \nabla \cdot \left(  \frac{\hbar^2}{2m(z)} \nabla \psi_{\mathbf{n},\lambda} \right) $ as follows:
\begin{align}
\label{eq: kin term}
   & - \nabla \cdot \left(  \frac{\hbar^2}{2m^*(z)} \nabla \psi_{\mathbf{n},\lambda}(\varX) \right) = \\
   & - \frac{\hbar^2}{2m^*(z)} \nabla^2 \psi_{\mathbf{n},\lambda}(\varX) - \frac{d}{dz} \left( \frac{\hbar^2}{2m^*(z)} \right) \pdv{\psi_{\mathbf{n},\lambda}}{z}
    \nonumber =  \\
   &  \frac{1}{L} e^{i (k_x x+k_y y)}
\chi_{n_x, n_y, \lambda} \nonumber \\
   & \left[ -  
   \frac{d}{dz} \left(
   \frac{\hbar^2}{2m^*(z)} \phi_{\mathbf{n},\lambda}^{'}(z)
   \right)
   + \frac{\hbar^2}{2m^*(z)} k_{n_x n_y}^2 \phi_{\mathbf{n},\lambda}(z)
   \right] \nonumber.
\end{align}
The constant spinor $\chi$ and the exponential appear in all terms in Eq. \eqref{eq: skyrme HF eqs}, thus they can be simplified and drop out of the the final equations \eqref{eq: skyrme final eqs}.

\subsection{Densities as a function of the orbitals}
\label{app: dens orbitals}
Number density, kinetic density and spin-orbit density may be computed from their definitions as functions of the occupied orbitals \cite{Schunck2019} applied to the wave functions \eqref{eq: orbitals}. Eqs. \eqref{eq: nabla psi} and \eqref{eq: laplacian psi} are also used to find
\begin{align}
    \rho(z) &= \sum_j \abs{\psi_j(\varX)}^2 = \, \frac{1}{L^2} \sum_{\mathbf{n},\lambda} \abs{\phi_{\mathbf{n},\lambda}(z)}^2 \\
    \tau(z) &= \sum_j \abs{\nabla \psi_j (\varX)}^2 \\ &= \frac{1}{L^2} \sum_{\mathbf{n},\lambda} \left( 
    \abs{\phi_{\mathbf{n},\lambda}'}^{2} + k_{n_x n_y}^2 \abs{\phi_{\mathbf{n},\lambda}}^2
    \right)   \nonumber \\
    \label{eq: spin orbit density final}
    J_z(z) &= \sum_j \psi_j^*(\varX)  \left( -i\right) \left( \nabla \cross \mathbf{\sigma} \right)_3 \psi_j(\varX) \\
    &= \sum_{\mathbf{n},\lambda} \psi_{\mathbf{n},\lambda}^*(\varX) K \psi_{\mathbf{n},\lambda}(\varX) \nonumber \\
    &= \frac{1}{L^2} \sum_{\mathbf{n},\lambda} \lambda k_{n_x n_y} \abs{\phi_{\mathbf{n},\lambda}(z)}^2 \nonumber
\end{align}
where only the z component of $\mathbf{J}$ does not vanish and Eq. \eqref{eq: helicity eigenstate} has been used.

\subsection{Hamiltonian in the plane waves basis}
\label{app: hamiltonian in plane waves}
We derive the Hamiltonian matrix in the plane waves basis $\left( \tilde{h}_{\mathbf{n},\lambda} \right)_{k,k'}$ (Eq. \eqref{eq: eig plane waves}). We start from the real space DFT equations \eqref{eq: skyrme final eqs} and Fourier-expand the orbitals as $\phi(z)= \frac{1}{\sqrt{L}} \sum_{k'} c_{k'} e^{ik'z}$. Then, we project on the $k$ plane wave by multiplying by $e^{-ikz}/\sqrt{L}$ and integrating over $z$ for $-L/2 \le z \le L/2$. 
The multiplicative terms are simple to treat and one easily finds the Fourier transform 
\begin{align}
    & \tilde{U}(k-k') = \frac{1}{L} \int_{-L/2}^{L/2} dz \, e^{-i(k-k')z} \\ 
    &\left( U(z) + v(z) + \lambda k_{n_x n_y} W(z) + \frac{\hbar^2}{2m^{*}(z)} k_{n_x n_y}^2 \right) \nonumber.
\end{align}
The derivative term is slightly more involved and is discussed in detail. 
We simplify the notation by defining $B(z) = \frac{\hbar^2}{2m^{*}(z)} $ and dropping the subscripts ${\mathbf{n},\lambda}$ and move on to compute
\begin{align}
    \frac{1}{\sqrt{L}}  \int_{-L/2}^{L/2}\,dz e^{-ikz} \bigg[ - \frac{d}{dz} \big( B(z)
    \phi'(z)
   \big)   \bigg].
\end{align}
An integration by parts, followed by inserting $\phi'(z)= \frac{i}{\sqrt{L}} \sum_{k'} k' c_{k'} e^{ik'z} $, gives
\begin{align}
    & \frac{1}{\sqrt{L}}  \int \, dz B(z)
    \phi'(z) \frac{d}{dz} e^{-ikz} = \\
    & -i \frac{k}{\sqrt{L}}  \int dz \, B(z)
    \phi'(z) e^{-ikz} = \nonumber \\
    & k \sum_{k'} k' c_{k'} \, \frac{1}{L} \int dz B(z) e^{-i(k-k')z} \nonumber = \\
    & k \sum_{k'} \tilde{B}(k-k') k' c_{k'} \nonumber
\end{align}
where 
\begin{align}
\tilde{B}(k-k') = \frac{1}{L} \int_{-L/2}^{L/2} dz\, e^{-i(k-k')z} \frac{\hbar^2}{2m^{*}(z)}.
\end{align}
In case effective mass terms are absent, $m^{*}(z)=m$, $\tilde{B}(k-k')$ is simply equal to $\frac{\hbar^2}{2m} \delta_{k,k'}$ and one recovers in $\tilde{h}_{k,k'}$ the usual kinetic term $\frac{\hbar^2}{2m} k^2$.
Summing the $\tilde{B}$ and  $\tilde{U}$ terms, one finds the Hamiltonian matrix
\begin{equation}
    \tilde{h}_{k,k'} = k \tilde{B}(k-k') k' + \tilde{U}(k-k').
\end{equation}

\section{Details on the static response theory}
\label{app: static response theory}
Further details on the static response theory are given in what follows, and in particular the key equation \eqref{eq: ev quadratic} is derived.
The starting point is the formula for the density fluctuation \eqref{eq: delta rho real space}, using which $\chi$ can be expressed as the functional derivative
\begin{align}
\label{eq: chi def}
    \chi(\varX,\varX') = \fdv{\rho_v(\varX)}{v(\varX')} \eval_{v=0}.
\end{align}
We now want to prove that the dependence of the energy on the perturbation is instead quadratic. Indeed, this can be verified by first expanding $E[v]$ (understood as a functional of $v$) around the unperturbed system $v=0$, namely \cite{SenatoreBook}
\begin{align}
\label{eq: expansion E[v] 1 }
    & E[v] - E[0] = \int d\varX \fdv{E}{v(\varX)}\eval_{v=0} v(\varX)  + 
    \\& \frac{1}{2} \int d\varX \int d\varX' \secondfdv{E}{v(\varX)}{v(\varX')}\eval_{v=0} v(\varX) v(\varX') \nonumber.
\end{align} 
Then, we notice that $\fdv{E}{v(\varX)}=\rho_v(\varX)$ as the external potential enters $E[v]$ the energy with the term $\int d\varX v(\varX) \rho(\varX)$ and thus $\fdv{E}{v(\varX)}\eval_{v=0}=\rho_0$. Differentiating the energy twice and inserting Eq. \eqref{eq: chi def}, moreover, we find
\begin{align}
\secondfdv{E[v]}{v(\varX)}{v(\varX')} = \fdv{\rho_v(\varX)}{v(\varX')} = \chi(\varX,\varX').
\end{align}
Therefore, Eq. \eqref{eq: expansion E[v] 1 } can be recast as \cite{SenatoreBook}
\begin{align}
\label{eq: delta E func v}
    & E[v] - E[0] = \int d\varX v(\varX) \rho_0 + \\& \frac{1}{2} \int d\varX \int d\varX' \chi(\varX,\varX') v(\varX) v(\varX') \nonumber,
\end{align}
and we immediately see that the first-order term vanishes, $v$ being periodic. (A more general argument is presented in Ref. \cite{giuliani_vignale_2005}).
We also remind that the homogeneous matter response depends only on $\varX-\varX'$ due to translational invariance, i.e $\chi(\varX,\varX')=\chi(\varX-\varX')$.

Then one can transform Eq. \eqref{eq: delta E func v} to momentum space inserting the Fourier expansions 
\begin{align}
    \label{eq: fourier exp}
    \delta \rho(\varX) &= \sum_\varK \rho_\varK e^{i\varK \cdot \varX}, \quad v(\varX) = \sum_\varK v_\varK e^{i\varK \cdot \varX}
    \\
    \label{eq: chi fourier transform}
    \chi(\varX - \varX') &=  \frac{1}{\Omega} \sum_\varK \chi(\varK) e^{i\varK \cdot (\varX- \varX') }.
\end{align}
Then
\begin{align}
\label{eq: delta E fourier}
    E[v] - E[0] =
    \frac{\Omega}{2} \sum_\varK v_\varK \chi(\varK) v_{-\varK}.
\end{align}
If the monochromatic potential \eqref{eq: periodic v} is considered in place of a generic perturbation, and if the relations $\rho_0=A/\Omega$ and $\chi=\chi(\abs{\varQ})$ that hold for uniform matter are employed, we find that the energy per particle of the perturbed system is given by \cite{SenatoreBook}
\begin{equation}
    \delta e_v = e_v - e_0 = \frac{\chi(q)}{\rho_0} v_q^2.    %+ C_4 v_q^4
\end{equation}

\section{EDF response in the thermodynamic limit}
\label{app: edf tl resp}
The dynamic response of a large class of generalized Skyrme EDFs has been determined in the thermodynamic limit analytically in Ref. \cite{PASTORE20151} and references therein. 
We summarize the main formulas here for the case of PNM and SNM. A slightly different notation is also introduced.

First, for later convenience we define $K_{bulk}$ as
\begin{align}
\label{eq: K bulk}
    K_{bulk} = \sum_\gamma c_\gamma \gamma (\gamma+1) \rho^{\gamma-1}.
\end{align}
Then, the following $W$ functions are defined as in Ref. \cite{PASTORE20151}, namely
\begin{align}
    W_1(q)/g & =  K_{bulk} - \left( 2 C^{\Delta\rho} + \frac{C^{\tau}}{2} \right) q^2 \\
    W_2/g & = C^{\tau} \\
    W_{so}/g &= C^{\nabla J} .
    %W_{so} &= 4 C_0^{\nabla J} \, (\rm{SNM}), \quad 2 C_{nm}^{\nabla J} \, (\rm{PNM}) .
\end{align}
$W_2$ is a constant proportional to $C^{\tau}$, while $W_1$ mixes the $C^{\tau}$ and $C^{\delta}$ coefficients and carries a momentum dependence through $q^2$. Lastly, $W_{so}$ is a spin-orbit constant.

Now, we introduce adimensional functions $X$ and insert them into $\chi(q)$ (eq. (67), Ref. \cite{PASTORE20151}). 
With $k=q/2q_F$, we define $\tilde{\rho}$ as $\rho$ in SNM and $2 \rho$ in PNM. With this trick, the expressions for SNM \cite{Pastore2012snm} and PNM \cite{Pastore2012pnm} are identical.
The $X$ functions are derived from the corresponding $W$ functions by means of
\begin{align}
    X_1 &= \frac{m^{*}c^2}{(\hbar c)^2} \tilde{\rho} \frac{W_1(q)}{q_F^2} \\
    X_2 & = \frac{m^{*}c^2}{(\hbar c)^2} \tilde{\rho} W_2 \\
    X_{so} &= \frac{m^{*}c^2}{(\hbar c)^2} \tilde{\rho} W_{so}.
\end{align}

We further elaborate on $X_1$ by splitting it as the sum or a bulk and a momentum-dependent contributions:
\begin{align}
    X_1(k) = X_{bulk} + X_{surf}(k)
\end{align}
with 
\begin{align}
    X_{bulk} &= g \frac{m^{*}c^2}{(\hbar c)^2}  \frac{\tilde{\rho}}{q_F^2} K_{bulk}     \\
    X_{surf}(k) &= -4g \left( 2 C^{\Delta\rho} + \frac{C^{\tau}}{2} \right) \frac{m^{*}c^2}{(\hbar c)^2} \tilde{\rho} k^2.
\end{align}
Finally, by using $\chi(q)=-\rho \, 2 m_{-1}(q)/A$, with $m_{-1}$ being the inverse energy-weighted sum rule of the strength function, and collecting some constant factors, one ends up with following formula for the TL response of a nuclear EDF:
\begin{align}
\label{eq: formula pastore}
    \chi(q) &= - 3 \frac{m^{*}c^2}{ (\hbar c)^2 } \frac{\rho}{q_F^2} f(k) \\ &\Bigg[
    \left( 1 + \frac{3}{8} X_2 \right)^2 
    + \frac{3}{4} \bigg( X_1(k) + X_2 (1-k^2) \bigg) f(k) \nonumber
    \\ & - \frac{3}{64} X_2^2 \left( 2 + \frac{26}{3}k^2 + (1-k^2) f(k) \right) f(k)
    \nonumber \\ & - \frac{3}{8} k^2 f(k) X_{so}^2 \bigg( 1 + 3(1-k^2) f(k) \bigg)
    \Bigg]^{-1} .
\end{align}
with $f(k)$ defined in Eq. \eqref{eq: fk lind}.

\bibliography{bibliography.bib} 

\end{document}